\documentstyle[eqsecnum,psfig,aps,12pt]{revtex}
\def\beq{\begin{equation}}
\def\eeq{\end{equation}}

\def\step{\\[-2ex]}

\def\bea{\arraycolsep .1em \begin{eqnarray}}
\def\eea{\end{eqnarray}}
\def\Tr{{\rm Tr}}

\def\dq{{q \llap{/}}\,}

\def\de{\delta}
\def\Ga{\Gamma}

\def\eq#1{(\ref{#1})}
\def\Eq#1{Eq.~(\ref{#1})}
\def\Es#1{Eqs.~(\ref{#1})}

\def\s0#1#2{\mbox{\small{$ \frac{#1}{#2} $}}}
\def\0#1#2{\frac{#1}{#2}}

\begin{document}
\begin{center}

\thispagestyle{empty}

{\normalsize\begin{flushright}
CERN-TH-2001-084\\[18ex] 
\end{flushright}}

\mbox{\large \bf Optimised Renormalisation Group Flows} \\[10ex]

{Daniel F. Litim
\footnote{Daniel.Litim@cern.ch}
}
\\
{\it Theory Division, CERN, CH -- 1211 Geneva 23.}
\\[10ex]
 
{\small \bf Abstract}\\[2ex]
\begin{minipage}{14cm}{\small 
    
    We study the optimisation of exact renormalisation group (ERG)
    flows.  We explain why the convergence of approximate solutions
    towards the physical theory is optimised by appropriate choices of
    the regularisation.  We consider specific optimised regulators for
    bosonic and fermionic fields and compare the optimised ERG flows
    with generic ones. This is done up to second order in the
    derivative expansion at both vanishing and non-vanishing
    temperature. We find that optimised flows at finite temperature
    factorise. This corresponds to the disentangling of thermal and
    quantum fluctuations.  A similar factorisation is found at second
    order in the derivative expansion.  The corresponding optimised
    flow for a ``proper-time renormalisation group'' is also provided
    to leading order in the derivative expansion.

}
\end{minipage}
\end{center}

\newpage
\pagestyle{plain}
\setcounter{page}{1}
\noindent 
%********|*********|*********|*********|*********|*********|*********|****
\section{Introduction}
%********|*********|*********|*********|*********|*********|*********|****

Wilsonian renormalisation group techniques \cite{Wilson,Wegner} like
the exact renormalisation group (ERG)
\cite{Polchinski,CW,Ellwanger:1994mw,Morris:1994qb} are important
tools for addressing non-perturbative problems within quantum field
theory (for recent reviews, see Refs.~\cite{Bagnuls:2000ae,Berges:2000ew}).
They are similar in spirit to the block-spin action invented in
condensed matter physics, and their particular strength is their
flexibility, allowing for systematic approximations without being tied
to the small coupling region. The ERG is based on an infrared (IR)
regularisation with momentum scale parameter $k$ of the full
propagator, which turns the corresponding effective action into a
scale dependent functional $\Gamma_k$. The ERG flow describes the
change of the effective action under an infinitesimal variation of the
IR scale $k$. It thereby interpolates between the initial UV action
$\Gamma_{k=\Lambda}$ and the full quantum effective action
$\Gamma\equiv\Gamma_{k=0}$. Although the flow depends explicitly on
the specific infrared regulator chosen, the endpoint of the integrated
full flow does not.  \step

An explicit computation of the IR effective theory based on the ERG
flow requires the specification of the field content, the initial
condition $\Gamma_\Lambda$ and the choice of a particular IR
regulator. The UV initial condition is typically given by the
classical action. Hence, the main physical information is contained in
the ERG flow itself. Most problems of physical interest are too
complex to be solved exactly and an application of this formalism --
as of any other method -- is bound to certain approximations.
Furthermore, the flow equation is equivalent to infinitely many
coupled partial differential equations, which would seem very
difficult to solve exactly. Therefore, one has to resort to some
approximations or truncations which allow, at least in principle, for
a systematic computation of the full quantum effective action. In
order to provide reliable physical predictions, like a high precision
computation of universal critical exponents, it is mandatory to
provide a good control for approximated ERG flows.  \step

A number of systematic expansion schemes for flow equations are known,
including standard perturbation theory. Non-perturbative expansions of
the effective action, not bound to the weakly coupled regime, are the
derivative expansion, expansions in powers of the fields, or
combinations thereof. For example, the leading order of the derivative
expansion retains only an effective potential and a standard kinetic
term, and contains non-perturbative information as it corresponds to
the resummation of infinitely many perturbative loop diagrams. The
study of approximate quantum effective actions along these lines is a
sensible procedure since the underlying expansions admit a systematic
improvement to higher order.  \step

Solutions to truncated flow equations display a spurious dependence on
the IR regulator
\cite{Ball:1995ji,Litim96,Aoki:1998um,LPS,Freire:2000sx,Sumi:2000xp,Litim:2000ci,Latorre:2000qc}.
This is similar to the scheme dependence of physical observables
observed within perturbative QCD \cite{Stevenson}, or the truncation
dependence of solutions to Schwinger-Dyson equations. Its origin is
the following. The IR regulator couples, through the flow equation, to
all vertex functions of the theory.  The flow trajectory of the
functional $\Gamma_k$ in the space of all effective action functionals
depends on the regulator.  Hence, the regulator -- while regulating
the flow -- also modifies the effective interactions at intermediate
scales $k\neq 0$.  In other words, the effective action at
intermediate scales still has some memory of the details of how the
integrating-out of degrees of freedom has been performed. This
regulator dependence is of no relevance for the full flow. Eventually,
the convergence towards the full quantum effective action for {\it
  any} regulator ensures that all regulator-induced interactions
cancel out in the physical limit.  Approximations imply that certain
vertex functions are neglected.  Then, not all regulator-induced
interactions cancel out for $k\to 0$: the missing back coupling of the
neglected vertex functions is responsible for regulator-dependent
terms in the physical limit. In consequence, approximations to the
full quantum effective action depend spuriously on the scheme.  \step

Recently, a new line of reasoning has been put forward which
essentially turns this observation around \cite{Litim:2000ci}: given
that the solution of a truncated flow depends on the regulator, it
should be possible to identify specific ones which ``optimise'' the
physical content of a given approximation. Optimised regulators
stabilise the flow and lead to a faster convergence of expansions, such
that the main physical information is almost exclusively contained
within a few leading terms, and higher order contributions remain
small \cite{Litim:2001fd}.  \step

In Ref.~\cite{Litim:2000ci}, we have derived a simple and generic
optimisation criterion for ERG flows, based only on the full inverse
propagator at vanishing field. Given the set of possible IR
regulators, the criterion allows to distinguish the quality of
regulators in the sense outlined above.  In the present paper we study
a specific ``optimised'' regulator for both bosonic and fermionic
degrees of freedom. To be more explicit, we introduce the
ERG flow for the effective action
\cite{CW,Ellwanger:1994mw,Morris:1994qb}.  For bosonic fields $\phi$,
it is given by
\beq\label{flow} 
\partial_t\Gamma_k[\phi]=
\frac{1}{2}\Tr 
\left( \frac{\de^2\Gamma_k}{\delta \phi(q)\delta \phi(-q)}+R_k
\right)^{-1} \partial_t R_k \,.      
\eeq
Here, the trace denotes a sum over all loop momenta and indices, and
$t=\ln k$ is the logarithmic scale parameter. The flow has a simple
one-loop structure. The Wilsonian ``integrating-out'' is achieved by
the infrared regulator $R_k$. It regulates the propagator for small
momenta, while the insertion $\partial_t R_k$ cuts off the
large-momentum contributions. In total, only a small momentum window
about $q^2\approx k^2$ contributes to the flow. Apart from a few
constraints displayed later, the function $R_k$ can be chosen at will.
A ``good'' choice for the regulator function is at the root for
reliable physical predictions, and we consider, for the bosonic
fields, the optimised regulator 
\beq
\label{Ropt1} 
R^{\rm  opt}_k(q^2) = Z_k\cdot (k^2-q^2)\Theta(k^2-q^2)\,, 
\eeq 
where $Z_k$ is an appropriately defined wave function renormalisation.
This regulator is particularly simple: for loop momenta $q^2> k^2$ it
vanishes identically and the effective propagator appearing in the
flow equation is not modified; for loop-momenta $q^2<k^2$ it acts like
a momentum-dependent mass term in such a way that the inverse
effective propagator $\sim q^2+R_k(q^2)$ becomes a momentum
independent constant. In consequence, the effective infra-red
propagator does no longer distinguish between the different modes with
$q^2<k^2$.  \step

Optimised flows based on \Eq{Ropt1} derive from a generic optimisation
criterion \cite{Litim:2000ci}, and have a number of remarkable
properties. The optimised flow leads to the fastest decoupling of
heavy modes, in accordance with the decoupling theorem
\cite{Appelquist:1975tg}. In the limit $k\to 0$, optimised flows
smoothly approach a convex effective action, owing to a simple
analytic pole of the flow \cite{Litim:2000ci,Litim:2001fd}. At non-vanishing
temperature, the optimised flow factorises: the contributions from
thermal and quantum fluctuations are disentangled, unlike for generic
flows. A similar factorisation of the flow holds to second order in
the derivative expansion for field-independent wave function
renormalisations, and a partial factorisation is found for the general
case. Finally, the optimised flow has a very simple analytic
structure. This facilitates their study and is helpful for both
analytical or numerical considerations. All these properties lead to a
stabilisation of the flow and an improved convergence towards the
physical theory.  Analogous results for fermionic flows are discussed
as well.  \step

We also study this question within an RG formalism based on a
proper-time regularisation of the operator trace for the one-loop
effective action \cite{Liao:1996fp}, which we call ``proper-time
renormalisation group'' (PTRG) for short. In contrast to the ERG, the
PTRG has no path integral derivation, which makes the conceptual
reasoning more difficult \cite{Litim:2001hk}. Still, owing to the
close similarity to the ERG at leading order in the derivative
expansion, it is possible to identify the analogue of \Eq{Ropt1} for
the PTRG.\step

The format of the paper is as follows. We introduce the physical ideas
behind the generic optimisation condition. Explicit realisations for
bosonic and fermionic degrees of freedom are introduced as well
(Section~\ref{Optimisation}). The main characteristics of optimised
flows are discussed to leading order in the derivative expansion, and
contrasted with those of generic flows (Section~\ref{Derivative}). We
then turn to the discussion of quantum field theories at finite
temperature. We show that optimised thermal flows factorise on the
level of the flow equation, unlike generic flows. A simple physical
explanation for the factorisation is provided (Section~\ref{Thermal}).
Next, we consider the extension to higher orders in the derivative
expansion. The cases of field dependent or independent wave-function
renormalisations are both discussed, and a similar factorisation for
optimised flows is established (Section~\ref{Anomalous}). Finally, we
provide the corresponding optimised proper-time cut-off for the PTRG
(Section~\ref{HeatKernel}). Due to the qualitative difference of the
topics studied, we discuss our findings separately at the end of the
corresponding sections. We close with a summary and an outlook
(Section~\ref{Discussion}). Three Appendices contain technical details
and explicit expressions for optimised flows.

%********|*********|*********|*********|*********|*********|*********|****
\section{Optimisation}\label{Optimisation}
%********|*********|*********|*********|*********|*********|*********|****

In this section, we discuss a generic optimisation
criterion for ERG flows for Euclidean quantum field theories. In
particular, we provide a simple and explicit optimised regulator for
both bosonic and fermionic flows. Prior to this, we have to review a
few basic properties of IR regulator functions, which are at the root
of the subsequent considerations.

%********|*********|*********|*********|*********|*********|*********|****
\subsection{Regulators}
%********|*********|*********|*********|*********|*********|*********|****

The flow equation \eq{flow} is defined through the infrared regulator
functions $R_k(q^2)$ and $R_{F,k}(q^2)$,
respectively \cite{CW,Ellwanger:1994mw,Morris:1994qb}. 
These operators depend on an
infrared scale $k$, which induces a scale dependence. When written in
terms of the scale-dependent effective action $\Ga_k$, the scale
dependence is given precisely by the flow equation \eq{flow}. The
right-hand side of \Eq{flow} contains the full inverse propagators and
the trace denotes a sum over all indices and integration over all
momenta.\step

The regulator scheme (RS) functions can be chosen at will, however,
within some basic restrictions. These restrictions ensure that the
flow equation is well-defined, thereby interpolating between an
initial action in the UV and the full quantum effective action in the
IR. More specifically, it is required that
\beq\label{I}
\lim_{q^2/k^2\to 0}R_k(q^2)>0\,.
\eeq 
This ensures that the effective propagator at vanishing field remains
finite in the infrared limit $q^2\to 0$, and no infrared divergences
are encountered in the presence of massless modes. This property makes
$R_k$ an infrared regulator. If the limit \eq{I} is finite, we call
the corresponding regulator {\it masslike}. 
The second requirement is the vanishing
of $R_k$ in the infrared,
\beq\label{II}
\lim_{k^2/q^2\to 0}R_k(q^2)\to 0\,. 
\eeq
This guarantees that the regulator function is removed in the physical
limit, where the scale-dependent effective action $\Gamma_k$ reduces
to the quantum effective action $\Ga=\lim_{k\to 0}\Ga_k$. The third
condition to be met is
\beq\label{III}
\lim_{k\to\Lambda}R_k(q^2)\to \infty\,.
\eeq
This way it is ensured that $\Ga_k$ approaches the microscopic action
$S=\lim_{k\to \Lambda}\Ga_k$ in the UV limit $k\to \Lambda$. 
In the rest of the paper, we put
$\Lambda=\infty$ for the UV scale, although our main line of reasoning
can be applied for finite $\Lambda$ as well. With this choice, the
regulator function depends only on $q^2$ and $k^2$, and it is convenient to
introduce a dimensionless function $r(q^2/k^2)$ as
\beq\label{rdef}
R_k(q^2)=Z_k\ q^2\ r(q^2/k^2)\,.
\eeq
with $Z_k$ an appropriate wave function renormalisation
(cf.~Section~\ref{Anomalous}); $Z_k\equiv 1$ 
to leading order in the derivative 
expansion.
Owing to the general conditions imposed on the regulator, the function
$r(y)$ ranges between $0\le r(y)\le\infty$. \step

Another condition concerns the proper normalisation of the regulator.
The normalisation fixes the scale at which the IR regulator
becomes effective. Let us employ the condition
\beq\label{cB_def}
R_k(q^2=c_B k^2)=Z_k\,c_B k^2
\eeq
for bosons (a similar condition holds for fermions, see below) and
$c_B>0$.\footnote{In Ref.~\cite{Litim:2000ci} the convention $c_B=1$ has
  been used.} The normalisation translates into the condition
$r(c_B)=1$. Two different choices for $c_B$ can always be mapped onto
each other through a rescaling of the IR scale $k$. Hence, a proper
normalisation is only of relevance for a comparison of different
regulators (as done in Ref.~\cite{Litim:2000ci}), or for theories
containing different bosonic and/or fermionic degrees of freedom,
where the {relative} normalisation of the regulators can become
important. \step

%********|*********|*********|*********|*********|*********|*********|****
\subsection{Optimisation criterion}
%********|*********|*********|*********|*********|*********|*********|****

Here, we discuss an optimisation criterion for ERG flows, which
ensures that flows like \Eq{flow} and approximations to it have good
convergence and stability properties. Following
Ref.~\cite{Litim:2000ci} (see also Ref.~\cite{Litim:2001fd}), we first
provide the general criterion for optimised choices of RS functions.
Then, more specifically, we apply this idea to bosonic and fermionic
theories with standard kinetic terms.\step

The physical information of the flow equation \eq{flow} is contained 
in the full effective inverse propagator, which is given by
\beq\label{Effective}
\0{\delta^{2}\Gamma_k[\phi]}{\delta \phi(q)\delta\phi(-q)}
+R_k(q^2)\,.
\eeq
Notice that \Eq{Effective} depends both on the fields
and on the RS function. The ERG flow is well-defined as long as the
full inverse propagator displays a gap,
\beq\label{Effective1}
\min_{q^2\ge 0}
\left(
\left.
\0{\delta^{2}\Gamma_k[\phi]}{\delta \phi(q)\delta\phi(-q)}
\right|_{\phi=\phi_0}
+R_k(q^2)\right) = C\, k^2 >0\,.
\eeq 
The functional derivative is evaluated at a properly chosen expansion
point $\phi_0$. The existence of the gap $C>0$ implies an IR
regularisation.  Furthermore, the gap is a prerequisite for the ERG
formalism.  Elsewise, \Eq{flow} becomes singular at points where the
full inverse effective propagator develops zero modes.\footnote{The
  case $C=0$ indicates that a saddle point expansion about $\phi_0$ is
  not applicable. Those points $\phi_0$ in field space with $C=0$
  correspond to an instability. The problem can be cured by chosing a
  more appropriate expansion point such that $C>0$. For related
  literature, see Ref.~\cite{Alexandre:1999ts}.} The size of the gap
$C$ in \Eq{Effective1} depends both on the RS function and on
dimensionless parameters like $\phi_0^2/k^2$ or mass ratios, specific
to the particular theory studied.\step
 
A natural optimisation criterion based on
\Eq{Effective1} consists of maximising the gap $C$ over the space of
all possible RS functions. Optimised RS functions are those for which
the maximum of $C$ is attained. The optimisation ensures that the
momentum-dependent kernel of the ERG flow is the most regular.
Therefore we expect that optimised flows are much more stable against
approximations and show better convergence properties.\step

The optimisation condition as
formulated above is, essentially, only sensitive to the momentum dependence
of the full inverse propagator. Dropping momentum-independent terms on
the left-hand side of \Eq{Effective1} changes the number $C$
accordingly, but leaves the explicit dependence on $R_k(q^2)$
unchanged. Therefore, the optimisation leads to the same set of
optimised RS functions as long as the implicit dependence of
$\Gamma^{(2)}_k[\phi]\equiv
\delta^{2}\Gamma_k[\phi]/\delta\phi(q)\delta\phi(-q)$
on the RS function remains negligible. For this reason, the
optimisation condition of Refs.~\cite{Litim:2000ci,Litim:2001fd} is based
only on the momentum-dependent terms of \Eq{Effective}. \step

From now on, we concentrate on a standard kinetic term.  The effect of
a field-dependent wave function renormalisation can be taken into
account as well (see Section~\ref{Anomalous} below). We expand the
full inverse propagator as $Z_k(q^2+Z_k^{-1}R_k(q^2)+\ldots)$ about
the regularised kinetic term. Finally, dropping the
momentum-independent terms transforms \Eq{Effective1} into
\beq\label{Effective2}
\min_{q^2\ge 0}
\left( q^2 +Z_k^{-1}\,R_k(q^2) \right) = C\, k^2 >0\,.
\eeq 
A far reaching consequence of the infrared regulator in
\Eq{Effective2} is the presence of a gap for all $k>0$, which follows
trivially from \Eq{I}.  The decisive difference between
\Eq{Effective1} and \Eq{Effective2} is that the size of the gap $C>0$
in \Eq{Effective2} depends only on the particular choice for the RS,
but not on the specific theory.  Rewriting \Eq{Effective2} in
dimensions of $k$ leads to
\beq
\label{P2}
P^2(y) 
\equiv      q^2/k^2  + R_k(q^2)/(Z_k\,k^2)
=       y[1+r(y)]\,,
\eeq
where $y\equiv q^2/k^2$. Expressed in
terms of \Eq{P2}, the size of the gap is given by
\beq
C=\min_{y\ge 0} P^2(y)\,.
\eeq
Any RS function is now characterised by the associated gap $C$.
The size of the gap can be made arbitrarily small. Effectively, this
corresponds to removing the IR regulator in the first place. However,
for fixed normalisation $c_B$, it cannot be made arbitrarily large,
$C<\infty$. Hence, the natural optimisation condition, which is
the requirement to maximise the gap, becomes
\beq
\label{Opt}
C_{\rm opt}=\max_{\rm (RS)}\left(\min_{y\ge 0}P^2(y)\right)\,.
\eeq
A few comments are in order. The maximum in \Eq{Opt} is taken over the
(infinite-dimensional) space of all possible RS functions. The number
$C_{\rm opt}$ is uniquely determined and reads $C_{\rm opt}= 2c_B$,
where $c_B$ is the normalisation of bosonic regulators. From now on,
we refer to \Eq{Opt} as an ``optimisation condition'', and all RS
functions for which $C=C_{\rm opt}$ are called solutions to the
optimisation condition. The space of solutions to the optimisation
condition is infinite-dimensional. Notice also that the
condition to minimise the gap is not an extremisation linked to the
regulator, because it corresponds to removing the IR regularisation. In
Ref.~\cite{Litim:2000ci}, a variety of different solutions have been
found, and some examples are given in Fig.~1 below. \step

In order to obtain \Eq{Effective2}, we have assumed a standard kinetic
term for the fields. Therefore, the resulting optimisation condition
\Eq{Opt} is independent on the specific theory. Once the
momentum-dependent part of $\Gamma^{(2)}_k$ depends on the fields, the
corresponding optimisation condition based on the momentum-dependent
part of \Eq{Effective1} is sensitive to the specific theory. Within a
derivative expansion, this happens starting from the second order
(cf.~the discussion in Section~\ref{Anomalous}). \step

The optimisation condition has a number of interpretations in more
physical terms (cf.~Refs.~\cite{Litim:2000ci,Litim:2001fd}).  It has been
shown that the radius of convergence for amplitude expansions is given
by $C$. Therefore the optimisation condition improves their
convergence. Furthermore, it leads to a smooth approach towards a
convex effective potential in the IR limit $k\to 0$. It has also been
shown that it improves the convergence of the derivative expansion
\cite{Litim:2001fd}. Finally, it is worth emphasizing that the optimisation
criterion is a rather mild condition: all regulator functions are
described by at most countably infinitely many parameters, because
$R_k$ is at least square integrable. Of these, only one parameter is
fixed by the optimisation criterion. \step

We now turn to the discussion of fermionic degrees of freedom $\psi$
and $\bar\psi$ \cite{Bornholdt:1993za,Jungnickel:1996fp}. The 
flow equation is given by
\beq\label{flowF} 
\partial_t\Gamma_k[\psi,\bar\psi]=
-\Tr 
\left( \frac{\de^2\Gamma_k}{\delta \psi(q) \delta \bar\psi(-q)} + R_{F,k}
\right)^{-1} \partial_t R_{F,k}\,.
\eeq
As usual, the trace sums over all loop momenta and indices. The
constraints on the function $R_{F,k}$ are similar to those on $R_k$
\cite{Jungnickel:1996fp}. Following Ref.~\cite{Jungnickel:1996fp}, we chose
the regulator proportional to $\dq$ and introduce
\beq
R_{F,k}(q)=Z_{F,k}\,\dq\ r_F(q^2/k^2)
\eeq
We chose the normalisation as
\beq\label{cF_def}
R^2_{F,k}(q^2=c_F k^2)=c_F k^2\,. 
\eeq
This translates into the condition $r_F(c_F)=1$. It has been shown that the 
fermionic analogue of the function \Eq{P2} 
is given by \cite{Jungnickel:1996fp}
\beq\label{P2F}
P_F^2(y) =y[1+r_F(y)]^2\,.
\eeq
Therefore, we can define the fermionic gap as
\beq
C_F=\min_{y\ge 0} P_F^2(y)\,,
\eeq
and the corresponding optimisation condition reads
\beq\label{OptF}
C_{F,\rm opt}=\max_{\rm (RS)}\left(\min_{y\ge 0}P_F^2(y)\right)\,.
\eeq
The optimised fermionic gap is uniquely determined through the
normalisation $c_F$ as $C_{F\rm opt}= 4c_F$. Conceptually, the
fermionic case is treated in the same way as the bosonic case.  The
sole difference stems from the fact that the bosonic kinetic term
contains two derivatives, while the fermionic kinetic term contains
only one.  Therefore, the functions \eq{P2} and \eq{P2F} entering the
optimisation condition are different.

%********|*********|*********|*********|*********|*********|*********|****
\subsection{Derivation of optimised bosonic and fermionic regulators}
%********|*********|*********|*********|*********|*********|*********|****

A lot of effort has been spent in order to provide explicit regulators
which lead to sufficiently simple and analytic ERG flows. For example,
the sharp cut-off provides a simple explicit flow
to leading order in the derivative expansion. For this reason, it is
one of the most intensively studied flows in the field
(cf.~Refs.~\cite{Wilson,Wegner,Hasenfratz:1986dm,Morris:1994qb,Tetradis:1996br,Comellas}).
Other attempts have been made based on power-like regulators $R_k\sim
q^2(k^2/q^2)^b$ for $b=1$ and $b=2$ \cite{Morris:1994ie}, or variants
of a mass-term regulator $R_k\sim k^2\Theta(k^2-q^2)$. These
regulators are still sufficiently simple from an algebraic point of
view, and lead to reasonably simple flows.\footnote{Of these, only the
power-like regulator with $b=2$ solves the optimisation condition
\Eq{Opt}.} However, in absence of an underlying ``guiding
principle'' it was not obvious how to make progress given the plethora
of possible regulators, and in particular, how to distinguish the
``quality'' of the corresponding flows. \step

Here, in turn, we take full advantage of the existence of a guideline
provided by the optimisation criterion. We propose a regulator which
(i) solves the optimisation criterion, (ii) is based on an additional
stability criterion for approximate flows, and (iii) leads to simple
explicit expressions for the corresponding flows. The heuristic
derivation runs as follows. The space of regulators which solve the
optimisation criterion is still infinite dimensional. Let us seek for
a ``simple'' solution to \Eq{Opt}. The simplest one corresponds to an
inverse propagator which is {\it flat}, {\it i.e.}
momentum-{independent}, $P^2\equiv C_{\rm opt}$. Take $C_{\rm opt}=1$.
This immediately implies, using \Es{rdef} and \eq{P2}, that
$R_k(q^2)=k^2-q^2$. Our naive Ansatz is consistent with \Es{I} and
\eq{III}, but {not} with the main requirement \Eq{II} for small
$k^2<q^2$. In order to fulfil \Eq{II}, the regulator has to be
cut-off above some loop momenta. Therefore, a natural proposal for the
bosonic case consists in taking
\beq\label{Ropt}
R_{k}^{\rm opt}(q^2)=(k^2-q^2)\Theta(k^2-q^2)\ .
\eeq
The ultraviolet modes $q^2> k^2$ are not touched by this regulator
because \Eq{Ropt} vanishes identically for $q^2>k^2$. In turn, for all
modes with $q^2\le k^2$ the regulator acts as a momentum-dependent
mass term $\sim (k^2-q^2)$ with the infrared limit $\sim k^2$ for
vanishing momenta. It is a masslike regulator. By
construction, the inverse propagator at vanishing field \Eq{Effective}
becomes momentum {in}dependent for all $q^2\le k^2$ (see Fig.~1). It
is this property which is responsible for the main characteristics of
the regulator: all infrared momentum modes below the scale $k$ are
treated in the same way since the effective inverse propagator does no
longer distinguish between them.\step

\begin{figure}[t]
\begin{center}
\unitlength0.001\hsize
\begin{picture}(500,550)
\put(180,450){\framebox{\large $P_{\rm opt}^2(q^2/k^2)$}}
\put(200,60){\large $q^2/k^2$}
\psfig{file=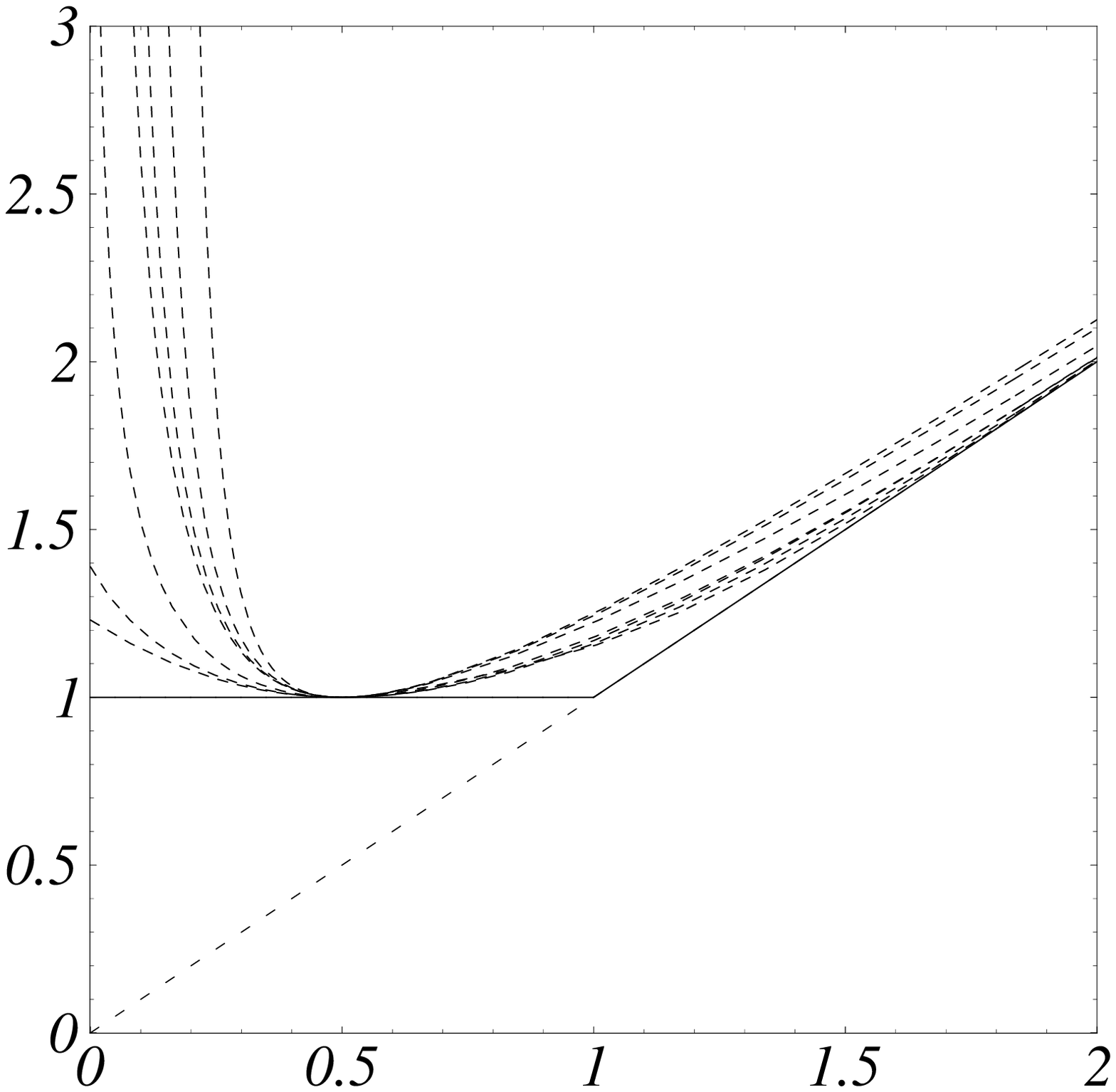,width=.45\hsize}
\end{picture}
\vskip-.5cm
\begin{minipage}{.92\hsize}
{\small {\bf Figure 1:} 
  Optimised inverse propagators $P_{\rm opt}^2$ for different
  regulators, normalised as $r(\s012)=1$. The regulator \Eq{Ropt} is
  given by the full line. The thin dashed line corresponds to $r=0$.
  All other dashed lines, given for comparison, correspond to the
  different optimised regulators of Fig.~3 in Ref.~\cite{Litim:2000ci}.  }
\end{minipage} 
\end{center}
\end{figure}

The dimensionless regulator function $r(y)$ is defined in \Eq{rdef}.
With the choice made in \Eq{Ropt} it follows that
\beq\label{ropt}
r_{\rm opt}(y)=\left(\frac{1}{y}-1\right)\Theta(1-y)\,.
\eeq
The regulator function is normalised with $c_B=\s012$. Such a 
normalisation can always be achieved. Other
normalisations are considered at the end of Section~\ref{Derivative}.
\step

In Fig.~1, we have displayed the effective inverse propagator at
vanishing field for different optimised regulators. The full line
represents \Eq{Ropt}, and the thin dashed line corresponds to $R=0$
(no regulator). The set of dashed lines correspond to the optimised
regulators discussed in Fig.~3 of Ref.~\cite{Litim:2000ci}. Here, they have
been given for comparison. Notice that all curves cross in the
normalisation point $r(c_B)=1$. All optimised propagators display
the same gap $C_{\rm opt}=2c_B$, but differ essentially in the curvature
around their minima.  \step

The fermionic analogue of \Eq{Ropt} is derived in essentially the same
way. Starting with \Eq{P2F}, imposing $P^2_F\equiv 1$ for small
momenta, and the general conditions \eq{I}, \eq{II} and \eq{III}, we
finally end up with
\beq
\label{RoptF}
R^{\rm opt}_{F,k}(q)=
\dq \left(\sqrt{\frac{k^2}{q^2}}-1\right)
\Theta(k^2-q^2)\,,
\eeq
normalised with $c_F=\s014$. In terms of a
dimensionless function $r_F(q^2/k^2)$, \Eq{RoptF} becomes
\beq\label{roptF}
r_{F, \rm opt}(y)=\left(\frac{1}{\sqrt{y}}-1\right)\Theta(1-y)\ .
\eeq
and $r_F(\s014)=1$. The non-analyticity of \Eq{roptF} is a direct
consequence of $R_{F,k}$ having only one mass dimension. We shall see
below that it is of no harm for the computation of fermionic flows
because \Eq{roptF} enters only in specific combinations such that the
non-analyticity disappears.

%********|*********|*********|*********|*********|*********|*********|****
\section{Derivative expansion}\label{Derivative}
%********|*********|*********|*********|*********|*********|*********|****

The flow equation \eq{flow} is a functional differential equation,
which, from a technical point of view, is equivalent to infinitely
many coupled partial differential equations for the couplings
parametrising the effective action $\Gamma_k$. A number of different
systematic approximation procedures for flows are known. In this
section, we consider flows to leading order in the derivative
expansion, based on expanding the operators of the effective action
according to the number of derivatives \cite{Golner:1986}. This leads
to a closed set of coupled partial differential equations for the
coefficient functions. We discuss the main structure of optimised flows
and contrast it with generic ones.\step

%********|*********|*********|*********|*********|*********|*********|****
\subsection{Specific flows}
%********|*********|*********|*********|*********|*********|*********|****

In order to make our subsequent reasoning more transparent, it is useful
to have an explicit example at hand. To that end, we consider an
$O(N)$-symmetric real scalar field theory in $d$ dimensions, the
linear sigma model. To leading order in the derivative expansion we
make the Ansatz \cite{Wetterich:1993be}
\beq\label{AnsatzGamma}
\Gamma_k=\int d^dx \left(U_k(\bar\rho) 
             + \012 Z_k(\bar\rho) \partial_\mu \phi^a\partial_\mu \phi_a
             + \014 Y_k(\bar\rho) \partial_\mu \bar\rho\partial_\mu \bar\rho
             +{\cal O}(\partial^4)
\right)
\eeq
for the effective action, with $\bar\rho=\s012 \phi^a\phi_a$. For
$N\neq 1$, there are two independent wave function factors $Z_k$ and
$Y_k$ beyond leading order in this expansion (cf.~Section~\ref{Anomalous}). 
To leading order in
the derivative expansion, the flow equation \eq{flow} reduces to a
flow for the effective potential, $\partial_t U_k$. The main physical
applications concern the non-trivial Wilson-Fisher fixed point in
$d=3$ and the computation of related universal quantities. \step

Inserting the Ansatz \eq{AnsatzGamma} into the basic flow equation,
and using $Z\equiv Y\equiv 1$, yields \cite{Wetterich:1993be}
\beq\label{FlowU}
\partial_t U_k= \012\int\0{d^dq}{(2\pi)^d}
\left(\0{(N-1)\partial_t R_k(q^2)}{q^2+R_k(q^2)+U'_k(\bar\rho)}
+\0{\partial_t R_k(q^2)}{q^2+R_k(q^2)+U'_k(\bar\rho)+2\bar\rho U_k''(\bar\rho)}\right).
\eeq
It is a second order non-linear partial differential equation. One
easily recognizes the contributions from the $N-1$ ``Goldstone'' modes
and the ``radial'' mode. A similar flow equation has been obtained for
the wave function renormalisations $Z_k$ and $Y_k$
\cite{Wetterich:1993be}. The momentum integration is regularised in
the UV, owing to the regulator term $\partial_t R_k(q^2)$ in the 
numerator, and in the IR due to $R_k(q^2)$ in the denominator.\step

%********|*********|*********|*********|*********|*********|*********|****
\subsection{Generic flows}
%********|*********|*********|*********|*********|*********|*********|****

For convenience we perform the angular part of the momentum
integration and rewrite the right-hand side of \Eq{FlowU} in terms of 
so-called threshold functions \cite{Wetterich:1993be} as
\beq\label{FlowUThreshold}
\partial_t U_k(\bar \rho)= 
2 v_d (N-1) k^d\ell^d_0\left(\0{U'_k(\bar\rho)}{k^2}\right)
+2 v_d      k^d\ell^d_0\left(\0{U'_k(\bar\rho)
                          +2\bar\rho U_k''(\bar\rho))}{k^2}\right)\,.
\eeq
The constants $v_d$ are given by
\beq\label{vd}
v_d^{-1}=2^{d+1}\pi^{d/2}\Gamma(\s0d2)\,,
\eeq 
and the functions $\ell^d_n(\omega)$ are defined as
\beq\label{ThresholdDef}
\ell^d_n(\omega)
=\left(\delta_{n,0}+n\right)
\int_0^\infty dy y^{\s0d2-1}\,\0{-y^2\, r'(y)}{(P^2(y)+w)^{n+1}}\ .
\eeq
While the flow \eq{FlowUThreshold} is specific for the theory defined
by \Eq{AnsatzGamma}, the functions \eq{ThresholdDef} are {\it not}.
These functions describe the generic structure of the flow to leading
order in the derivative expansion. The flows for different indices
$n>0$ are related by
\beq\label{HigherThreshold}
\partial_\omega \ell^d_n(\omega)=
-(n+\de_{n,0})\ell^d_{n+1}(\omega)\,.
\eeq
Therefore, it suffices to study the flows $\ell^d_0(\omega)$. \step

The fermionic analogue of the flow \eq{ThresholdDef} is
\cite{Jungnickel:1996fp}
\beq\label{ThresholdDefF}
\ell^d_{F,n}(\omega)
=\left(\delta_{n,0}+n\right)
\int_0^\infty dy y^{d/2}\,\0{-2y(1+r_F)r_F'}{(P_F^2(y)+w)^{n+1}}\ .
\eeq
and \Eq{HigherThreshold} holds equally for \Eq{ThresholdDefF}. Notice
the additional factor $1-r_F$ in the integrand, which arises due to
the Dirac structure of \Eq{RoptF}. We have used the normalisation
condition $r_F(\s014)=1$.  \step

It is evident that the characteristics of the flow, determined by the
choice of $R_k$, are entirely encoded within the functions
\Es{ThresholdDef} and \eq{ThresholdDefF} (or similar functions to
higher order in the derivative expansion). For a generic regulator,
these are complicated functions of the fields, which can be computed
explicitly only for very specific choices for the regulator. \step

Two properties of generic flows given in terms of $\ell^d_{0}(\omega)$
and $\ell^d_{F,0}(\omega)$ are worth being mentioned. First of all,
from their very definition and the constraints imposed on the
regulator function, we conclude that {any} function
\eq{ThresholdDef} for $n=0$ decays {at most} as $1/\omega$ for
$\omega\to\infty$ \cite{Wetterich:1993be}. Therefore, they describe
the decoupling of ``heavy'' modes from the flow, which is a
manifestation of the decoupling theorem \cite{Appelquist:1975tg}.
Secondly, all flows have a pole in $C+\omega$, where $C$ denotes the
gap. Both the analytical structure and the strength of the pole depend
on the regulator. From the general requirements for regulators, and
the explicit form of \eq{ThresholdDef}, it follows that the pole for
$n=0$ cannot be stronger than a simple analytical pole $\sim
1/(C+\omega)$. The pole of threshold functions has important physical
implications. It determines the approach to a convex effective
potential for theories within a phase of spontaneous symmetry breaking
\cite{Tetradis:1992qt,Berges:2000ew}.

%********|*********|*********|*********|*********|*********|*********|****
\subsection{Optimised flows}
%********|*********|*********|*********|*********|*********|*********|****

Now we turn our attention to the optimised regulators introduced in
\Es{Ropt} and \eq{RoptF}. The evaluation of \Eq{ThresholdDef} is
particularly simple because the $\Theta$-function cuts off the
momentum integration. Using \Eq{ropt}, \Eq{ThresholdDef} reduces to
two terms,
\beq\label{Threshold1}
\ell^d_0(\omega)
=\0{1}{1+w}\int_0^1 dy y^{\s0d2-1}
+\int_0^\infty dy y^{\s0d2-1}\,\0{(y-1)\delta(1-y)}{1-y\Theta(1-y)+w}\ .
\eeq
In the first term, the momentum integration is cut-off above $y\le 1$.
The function $P^2(y)$ remains a constant in this momentum regime,
which allowed to move the $\omega$-dependent term in front of the
momentum integration. The integrand of the second term contains
products of distributions. Since the integrand is proportional to
$\sim (1-y)\delta(1-y)$ the second term vanishes identically,
independently of the specific implementation for the
$\Theta$-function. The remaining momentum integration of the first
term becomes trivial and gives
\beq\label{ld0}
\ell^d_0(\omega)=
\0{2}{d}
\01{1+\omega}\,.
\eeq
We used the normalisation $r(\s012)=1$ and hence $P^2=1$ for $y\le 1$.
\step

For fermionic flows \eq{ThresholdDefF} and the regulator \eq{RoptF}, 
we find
\beq\label{ThresholdF1}
\ell^d_{F,0}(\omega)
=\01{1+\omega}\int_0^1 dy y^{d/2-1}
+2\int_0^\infty dy y^{\s0d2+1}\,
\0{[1+(\s01{\sqrt{y}}-1)\Theta(1-y)]
 (\s01{\sqrt{y}}-1)\delta(1-y)}{[\sqrt{y}+(1-\sqrt{y})\Theta(1-y)]^2+w}\ .
\eeq
The first term has a restricted momentum integration due to the
cut-off provided by the $\Theta$-function. The second term is more
involved, and the integrand even contains products of distributions.
Notice, however, that it contains the factor $\sim
(\s01{\sqrt{y}}-1)\delta(1-y)$ which is proportional to
$\sim(y-1)\delta(1-y)$. Therefore, the second term vanishes
identically and independent of the parametrisation of the
distributions and their products. The evaluation of the first term
gives finally
\beq\label{ld0F}
\ell^d_{F,0}(\omega)=\0{2}{d} \01{1+\omega}\,,
\eeq
and is identical to the bosonic flow. \step

%********|*********|*********|*********|*********|*********|*********|****
\subsection{Discussion}
%********|*********|*********|*********|*********|*********|*********|****

The flows described by the functions \eq{ld0} and \eq{ld0F} have the
simplest asymptotic structure for $\omega\to\infty$. This implies that
heavy modes decouple ``the fastest'' from the flow for optimised
regulators. For comparison, the sharp cut-off leads only to a
logarithmic decoupling $\sim \ln \omega$. Also, the decoupling does
not depend on the particular theory studied ({\it i.e.}~the
dimension), unlike the case for polynomial regulators $R_k\sim
q^2(k^2/q^2)^b$. Furthermore, the flow described by the functions
\eq{ld0} and \eq{ld0F} have the simplest and strongest pole structure
for $C+\omega\to 0^+$. The pole is a simple {analytic} one, which is
not the case for generic regulator functions. An immediate implication
of this structure is that the optimised flows \eq{ld0} and \eq{ld0F}
lead to a logarithmically smooth approach towards a convex effective
potential. This is very different from the sharp cut-off case, where
the approach is only exponentially \cite{Berges:2000ew}. A detailed
presentation of these results is given elsewhere.  \step

For completeness we quote the results for the flows \eq{ld0} and
\eq{ld0F} for arbitrary normalisation. While the normalisation is of
no relevance for a theory containing only bosonic or fermionic degrees
of freedom, their {relative} normalisation can become important for
theories containing bosons and fermions.  The normalisation conditions
\eq{cB_def} and \eq{cF_def} correspond to $r(c_B)=1$ and $r_F(c_F)=1$,
which can always be imposed because the functions $r(y)$ and $r_F(y)$
range between $0\le r,r_F\le \infty$.  The optimised gaps are $C_{\rm
  opt}= 2c_B$ and $C_{F\rm opt}= 4c_F$. For arbitrary $c_B$ the flow
\Eq{ld0} is obtained as
\beq\label{ld0c}
\ell^d_0(\omega)=
\0{2}{d}
\0{(2c_B)^{\s0d2+1}}{2c_B+\omega}
\eeq
In the fermionic case we find
\beq\label{ld0Fc}
\ell^d_{F,0}(\omega)=
\0{2}{d}
\0{(4c_F)^{\s0d2+1}}{4c_F+\omega}\,
\eeq
for the rescaled analogue of \Eq{ld0F}.\step

Finally, we note that flows $\ell^d_0(\omega)\sim \s01{1+\omega}$ have
been used earlier in the literature
\cite{Tetradis:1994ts,Litim:1995ex,Bergerhoff:1996zq}, however,
{without} the explicit knowledge of the corresponding regulator
function. These trial functions are sufficiently simple to allow for
analytical considerations. The motivation for their use was based on
the observation that the generic threshold function \Eq{ThresholdDef}
decays at most as $\omega^{-1}$ for large $\omega$. This suggested
that a regulator may exist which leads to $\ell^d_{0}(\omega)=A_d
(C+\omega)^{-1}$. Let us show how the normalisation $A_d$ can be fixed
from consistency arguments. We use the universal relation
$\ell^{2n}_{n}(0)=1$ \cite{Wetterich:1993be}, which holds for $d=2n$
dimensions, to identify the prefactor as $A_{2n}=\s01n C^{n+1}$. The
analytic continuation to arbitrary dimensions leads finally to our
results \eq{ld0c} and \eq{ld0Fc}. This reasoning shows that the Ansatz
$\ell^{d}_{0}(\omega)=\s02d C^{d/2+1}(C+\omega)^{-1}$ is
self-consistent.  However, we rush to add that these consistency
arguments are {necessary} conditions, but not {sufficient} ones:
{only} the explicit form of the regulator -- as given by \Es{Ropt} and
\eq{RoptF} -- finally justifies the few earlier computations. In
addition, \Es{Ropt} and \eq{RoptF} are explicitly required for the
computation of the flow at finite temperature 
(see Section~\ref{Thermal}) or to higher order 
(see Section~\ref{Anomalous}).

%********|*********|*********|*********|*********|*********|*********|****
\section{Thermal fluctuations}\label{Thermal}
%********|*********|*********|*********|*********|*********|*********|****

In this section we apply our reasoning in the context of a quantum
field theory coupled to a heat bath at temperature $T$, and to leading
order in the derivative expansion. We show that optimised flows, as
opposed to generic ones, disentangle the different contributions
related to thermal and quantum fluctuations, respectively. These
properties are realised, on the level of the flow equation, in terms
of an important {\it factorisation}. This leads to better convergence
properties of the flow itself. Approximate solutions of the flow
correspond to better approximations of the physical theory. \step

%********|*********|*********|*********|*********|*********|*********|****
\subsection{Imaginary time formalism}
%********|*********|*********|*********|*********|*********|*********|****

To be explicit, we consider a bosonic or fermionic field theory at
thermal equilibrium at the temperature $T$ within the Matsubara
formalism. This implies that periodic (resp.~antiperiodic) boundary
conditions for the bosonic (resp.~fermionic) fields are employed. As a
consequence, the $q_0$ integration in the flow equation \eq{flow} is
replaced by a sum over Matsubara modes $m=0,\pm 1,\pm 2,\ldots $. The
trace in \Eq{flow} contains a momentum integration, which is then
substituted as
\beq
\int \0{d^dq}{(2\pi)^d}\to T\sum_m \int \0{d^{d-1}q}{(2\pi)^{d-1}}\ .
\eeq 
In the integrand of \Eq{flow} the $q_0$ variable is replaced by
\beq\label{replace}
 q_0\to 2\pi         c_m T      
\eeq
where 
\bea
\label{cmb}
c_m&=&m\quad\quad\quad {\rm for\ bosons}\\
\label{cmf}
c_m&=&m+\s012\quad {\rm for\ fermions}\ .
\eea 
It is also useful to introduce the variable
\beq
\tau = 2\pi T/k
\eeq
for the following considerations. The replacement \eq{replace} implies
that functions $\ell^d_n(\omega)$ turn into temperature dependent
functions $\ell^d_n(\omega,\tau)$. We show that this function
factorises for the regulators \eq{Ropt} and \eq{RoptF}.

%********|*********|*********|*********|*********|*********|*********|****
\subsection{Dimensional reduction and fermion decoupling}
%********|*********|*********|*********|*********|*********|*********|****

Let us review a few basic facts known for generic flows at finite
temperature within the imaginary time formalism
\cite{Wetterich:1993be,Tetradis:1993xd,Jungnickel:1996fp,Litim98}.
\step

Bosonic fields within the Matsubara formalism display the phenomenon
of dimensional reduction at high temperature. This means that for $T$
large enough all non-vanishing Matsubara modes are suppressed due to
effective masses $\sim mT$ for the Matsubara modes with $m\neq 0$.
Only the $m=0$ mode survives in this limit, leading to an effective
theory in $(d-1)$-dimensions. For a generic bosonic regulator, the
finite temperature flow is given as
\beq\label{ThresholdT}
\ell^d_0(\omega,\tau)=
\0{v_{d-1}}{v_d}\0{\tau}{2\pi}\sum_m 
\int_0^\infty dy y^{\s0{d-1}{2}-1}
\0{-(y+c_m^2\tau^2)^2\, r'(y+c_m^2\tau^2)}{P^2(y+c_m^2\tau^2)+\omega}\,.
\eeq
The function $P^2$ is defined in \Eq{P2}. The asymptotic regime where
only the $m=0$ Matsubara mode contributes is reached for $\tau\to
\infty$. From \Es{ThresholdT} and \eq{HigherThreshold}, we deduce
\beq\label{Reduction1g}
\ell^d_n(\omega,\tau\to \infty)
=\0{v_{d-1}}{v_d}\0Tk \ell^{d-1}_n(\omega)\,. 
\eeq
On the other hand, the limit $\tau\to 0$ eventually switches on all
higher order Matsubara modes. It is straightforward to verify that
\beq\label{Reduction2g}
\ell^d_n(\omega,\tau\to 0)=\ell^d_n(\omega)\,. 
\eeq
The asymptotic limits for $T\to \infty$, \Eq{Reduction1g}, and $T\to 0$,
\Eq{Reduction2g}, display dimensional reduction for bosons as a function
of temperature for generic regulator function
\cite{Tetradis:1993xd}.\step

Fermions at finite temperature within the Masubara formalism can be
treated in essentially the same way. However, they happen to have no
$m=0$ mode as antiperiodic boundary conditions have to be used on the
$q_0$-integration. Hence, fermions do not display dimension reduction.
Rather, they decouple completely from the RG flow once the smallest
Matsubara mode is larger than the scale $k$. These properties can be
read-off from the temperature-dependent flow. For a generic fermionic
regulator, the flow $\ell^d_{F,0}(\omega)$ at finite temperature is
defined as
\beq\label{ThresholdTF}
\ell^d_{F,0}(\omega,\tau)=
\0{v_{d-1}}{v_d}\0{\tau}{2\pi}\sum_m 
\int_0^\infty dy y^{\s0{d-1}{2}-1}
\0{-2(y+c^2_m\tau^2)^2\, r_F'(y+c^2_m\tau^2)[1+r_F(y+c^2_m\tau^2)]}{P^2_F(y+c^2_m\tau^2)+\omega}\,.
\eeq
The function $P_F^2$ is given in \Eq{P2F}. The asymptotic regime where
the fermions decouple completely is reached for $\tau\to \infty$. From
\Eq{ThresholdTF}, we deduce that
\beq\label{Reduction1Fg}
\ell^d_{F,n}(\omega,\tau\to \infty)=0\,. 
\eeq
Again, the limit $\tau\to 0$ eventually switches on all higher order
Matsubara modes such that
\beq\label{Reduction2Fg}
\ell^d_{F,n}(\omega,\tau\to 0)=\ell^d_{F,n}(\omega)\,. 
\eeq
The asymptotic limits \eq{Reduction1Fg} and \eq{Reduction2Fg} describe
the decoupling of fermions in the high temperature limit for arbitrary
dimension and generic regulator function.

%********|*********|*********|*********|*********|*********|*********|****
\subsection{Optimised thermal flows and factorisation}
%********|*********|*********|*********|*********|*********|*********|****

We now turn to the optimised regulators \eq{Ropt} and \eq{RoptF}. For
this case, the flow \eq{ThresholdT} can be computed explicitly.
Inserting \Eq{Ropt} into \Eq{ThresholdT}, and following a reasoning
analogous to the one after \Eq{Threshold1}, we find
\beq\label{ldnT}
\ell^d_n(\omega,\tau)= B_d(\tau) \ell^d_n(\omega)
\eeq
with the temperature dependent function
\beq\label{Bd}
B_d(\tau)=
\0{d}{d-1}
\0{v_{d-1}}{v_d}  
\0{\tau}{2\pi}
\sum_m \left(1-c_m^2\tau^2\right)^{\s0{d-1}{2}}
\Theta\left(1-c_m^2\tau^2\right)\ .
\eeq
Notice that the temperature effects have {factorised}. This implies
that temperature cuts off all amplitudes $\omega$ in the same manner.
This is not the case for a generic regulator.  
%\step

Let us discuss the thermal threshold factor $B_d(\tau)$. In Fig.~2 the
thermal threshold factor $B_d(\tau)$ is displayed for $d=2, 3$ and $4$
dimensions. Every single Matsubara mode contributes to \Eq{Bd}
proportional to
\beq\label{single}
\sim \tau \left(1-c_m^2\tau^2\right)^{\s0{d-1}{2}}
\Theta\left(1-c_m^2\tau^2\right)\ .
\eeq
The $\Theta$-function is a remnant of the regulator \eq{ropt} and cuts
the $m$-th Matsubara mode off as soon as $k<c_m T/2\pi$. The factor
$\tau$ stems from the $q_0$-integration and the factor
$(1-c_m^2\tau^2)^{(d-1)/2}$ from the $d-1$ dimensional integration
over spatial loop momenta $|{\bf q}|$. These functions vanish outside
the interval $0\le\tau\le 1/c_m$. At the upper end they behave like
$(1/c_m-\tau)^{(d-1)/2}$ and vanish linearly with $\tau$ at the lower
end. This structure explains the spikes observed in Fig.~2, which are
located precisely at the points $\tau=1/c_m$ and due to the decoupling
of the $\pm c_m$-th Matsubara modes. Indeed, for $\tau>1$ only the
$m=0$ Matsubara mode yields a contribution to $B_d(\tau)$ in \Eq{Bd}.
The asymptotic regime where only the $m=0$ Matsubara mode contributes
is reached already for $\tau>1$ with $B_d(\tau\ge 1)=\tau d
v_{d-1}/2\pi v_d (d-1)$, or
\beq\label{Reduction1}
\ell^d_n(\omega,\tau\ge 1)=\0{v_{d-1}}{v_d}\0Tk \ell^{d-1}_n(\omega)\,. 
\eeq
Notice the difference to \Eq{Reduction1g}. Decreasing $\tau$ below
$\tau=1/c_m$ eventually switches on the $\pm c_m$ Matsubara modes. For
$\tau$ close to the points $1/c_m$, the term \eq{single} increases as
$(1/c_m-\tau)^{(d-1)/2}$ for decreasing $\tau$. This power law
explains why the spikes are more pronounced in lower
dimensions.\footnote{To higher order in the derivative expansion, the
spikes are smoothed-out for non-trivial wave function
renormalisation, cf.~Sect.\ref{Opt-2nd}.} In the limit $\tau\to 0$
it is straightforward to verify that $B_d(\tau\to 0)\to 1$ which
implies
\beq\label{Reduction2}
\ell^d_n(\omega,\tau\to 0)=\ell^d_n(\omega)\,. 
\eeq
This asymptotic limit is the same as \Eq{Reduction2g}.

\begin{figure}
\begin{center}
\unitlength0.001\hsize
\begin{picture}(500,550)
\put(80,450){\framebox{\large $B_d(\tau)$}}
\put(160,60){\large $\tau=2\pi T/k$}
\psfig{file=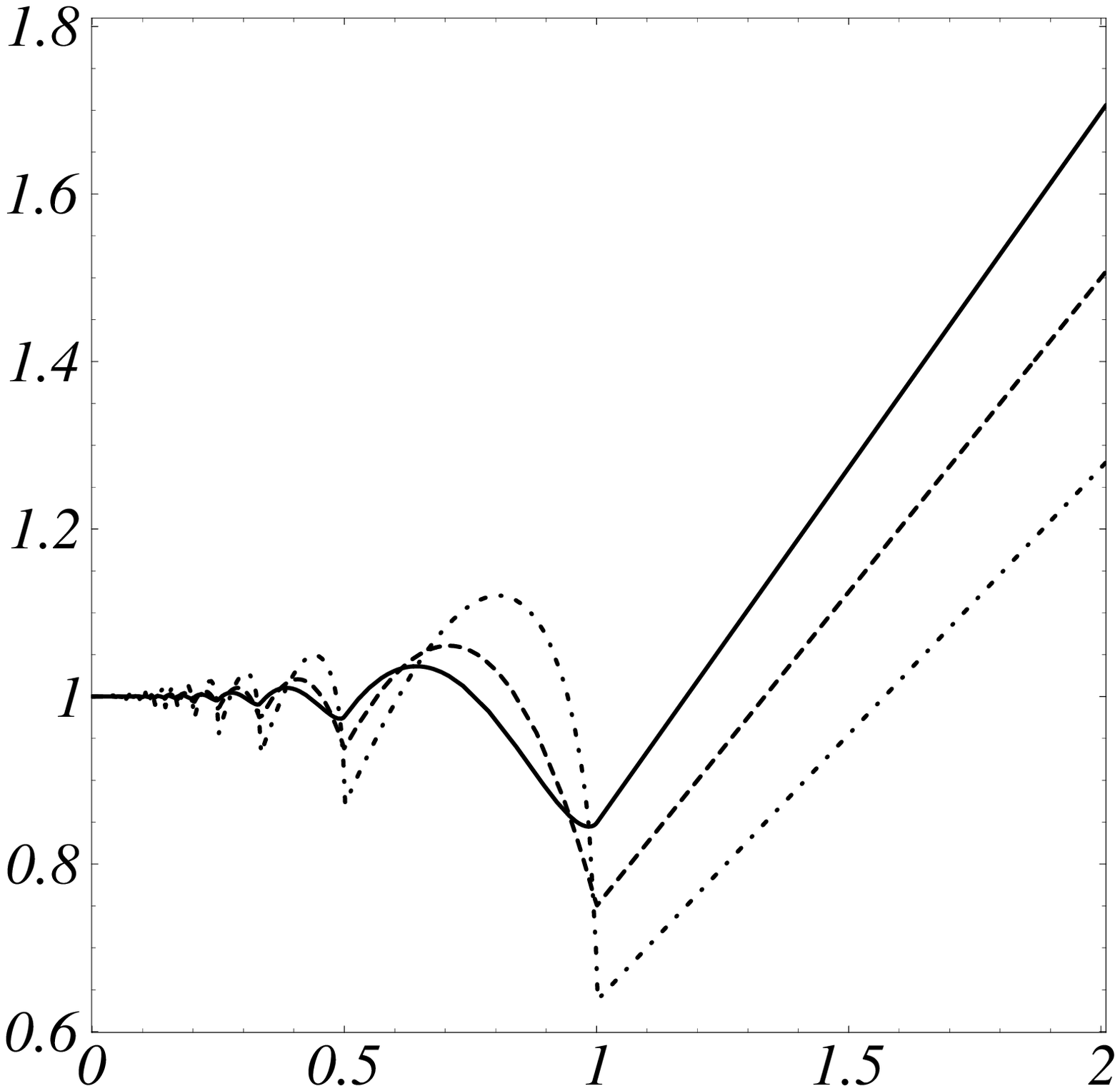,width=.45\hsize}
\end{picture}
\vskip-.5cm
\begin{minipage}{.92\hsize}
{\small {\bf Figure 2:} 
    Dimensional reduction for bosons, described by the bosonic thermal
  function $B_d(\tau)$ defined in \Eq{Bd}. $d=4$: full line, $d=3$:
  dashed line, $d=2$: dashed-dotted line. }
\end{minipage} 
\end{center}
\end{figure}

Turning to the optimal fermionic regulator \eq{roptF}, the momentum
integration in \Eq{ThresholdTF} can be performed explicitly to give
\beq\label{ldnTF}
\ell^d_{F,n}(\omega,\tau)= F_d(\tau) \ell^d_{F,n}(\omega)\ .
\eeq
As in the bosonic case, the temperature effects factorise from the
threshold effects. The fermionic thermal threshold factor $F_d(\tau)$
is given by
\beq\label{Fd}
F_d(\tau)=
\0{d}{d-1}
\0{v_{d-1}}{v_d}  
\0{\tau}{2\pi}
\sum_m \left(1-c_m^2\tau^2\right)^{\s0{d-1}{2}}
\Theta\left(1-c_m^2\tau^2\right)
\eeq
\Eq{Fd} is identical to its bosonic counterpart \eq{Bd} except for
the Matsubara sum which runs over $c_m=\pm\s012,\pm\s032,\ldots$ in
\Eq{Fd}. In Fig.~3 we have displayed the function $F_d(\tau)$ for
$d=2,3$ and $4$ dimensions. Again, the spikes have the same origin as
in the bosonic case and the same reasoning applies. The high
temperature limit at which the fermions decouple completely, is
already reached for $k\le \pi T$,
\beq
\ell^d_{F,n}(\omega,\tau\ge 2)=0\,.
\eeq
Notice the important difference to \Eq{Reduction1Fg}, where the
decoupling of fermions is only asymptotic. The limit $\tau\to 0$ is
equivalent to \Eq{Reduction2Fg}.
%\step
%\noindent

\begin{figure}[t]
\begin{center}
\unitlength0.001\hsize
\begin{picture}(500,550)
\put(300,450){\framebox{\large $F_d(\tau)$}}
\put(160,60){\large $\tau=2\pi T/k$}
\psfig{file=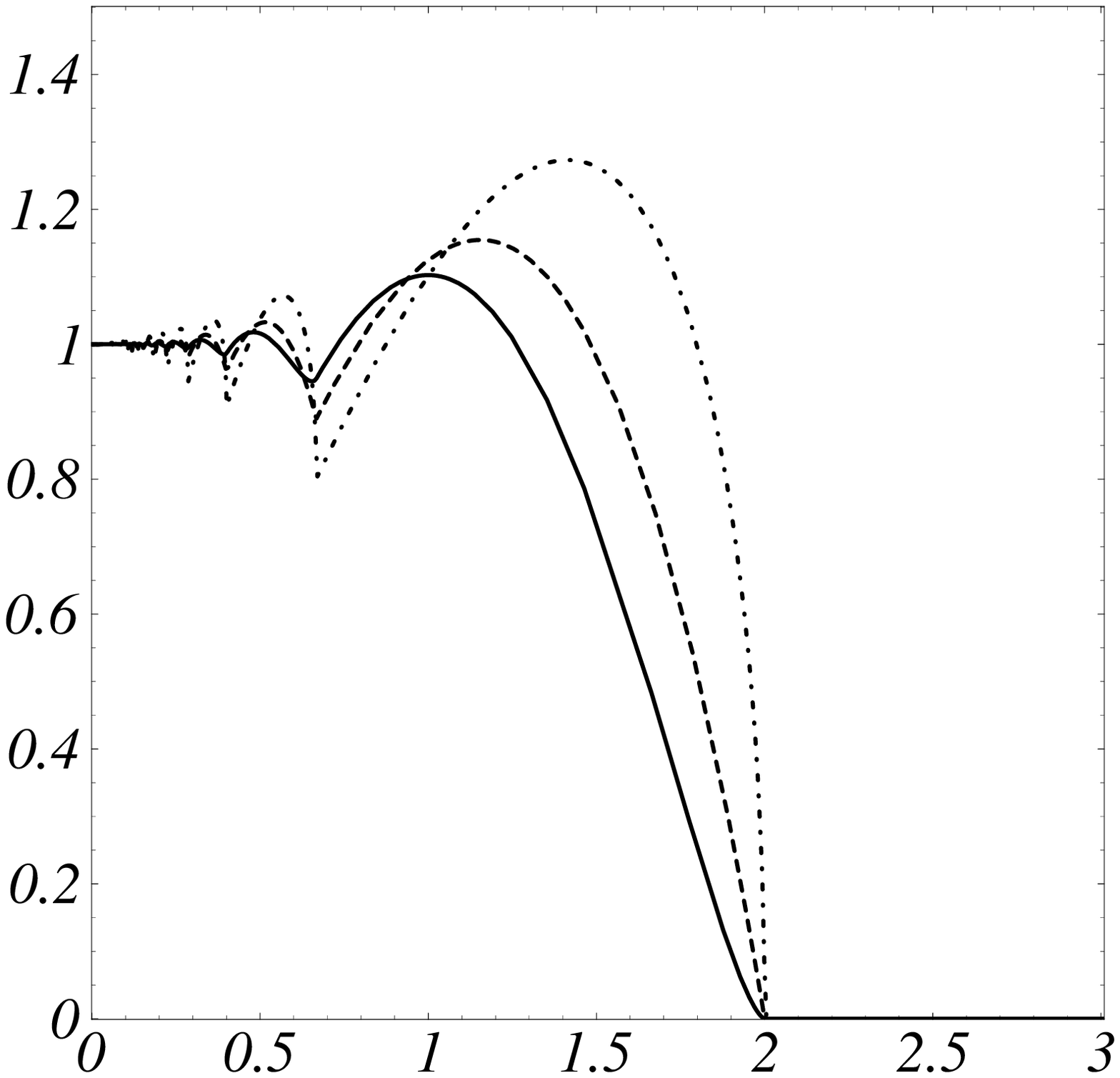,width=.45\hsize}
\end{picture}
\vskip-.5cm
\begin{minipage}{.92\hsize}
{\small {\bf Figure 3:} 
  Fermion decoupling at finite temperature described by the fermionic
  thermal function $F_d(\tau)$ given by \Eq{Fd}. $d=4$: full line,
  $d=3$: dashed line, $d=2$: dashed-dotted line.}
\end{minipage} 
\end{center}
\end{figure}

%********|*********|*********|*********|*********|*********|*********|****
\subsection{Discussion}
%********|*********|*********|*********|*********|*********|*********|****

The optimised regulators \eq{Ropt} and \eq{RoptF} correctly describe
dimensional reduction and fermion decoupling. In addition, they lead
to a thermal factorisation of the flow as observed in \Es{ldnT} and
\eq{ldnTF}. From a physical point of view, this fact is easily
understood. The imaginary time formalism compactifies the time
direction and the temperature modifies the {\it temporal} momentum
modes of the fields. The corresponding Matsubara mode, when compared
to the infrared scale $k$, leads to a {thermal} decoupling. To leading
order in the derivative expansion, the optimised regulator makes the
temperature blind for the quantum fluctuations. It cannot distinguish
between amplitudes of constant fields.  In turn, the quantum
fluctuations are sensitive to the field amplitudes, which are
responsible for the {mass} decoupling, similar to the case for
vanishing temperature. Clearly, these two effects have different
physical origins. As a consequence, it is natural to employ a
regulator which reflects this factorisation on the level of the flow
equation. \step

For a generic regulator function, the flows \eq{ThresholdT} and
\eq{ThresholdTF} are complicated functions of both the temperature and
the field amplitudes. They reflect dimensional reduction and fermion
decoupling. Typically, however, they do {not} factorise. This simply
means that a generic ERG flow entangles thermal and quantum
fluctuations even to leading order in the derivative expansion. This
is a direct consequence of the regulator term, whose coupling to the
different operators in the effective action is leading to a
field-dependent thermal decoupling of the different modes on the level
of the flow equation. This entanglement is of no relevance if the flow
can be solved {exactly}. In turn, for an {approximate} solution of
the flow, the factorisation on the level of the flow equation is most
helpful. It avoids a mixing of thermal and quantum fluctuations in a
regime where they can be disentangled, thereby minimising possible
artifacts due to the specific regulator function. As a consequence,
the flow itself is stabilised, and expansions of the flow show much
better convergence behaviour towards the physical theory. More
generally, it is expected that this line of reasoning applies for
generic optimised regulators.\step

Finally, the factorisation is very helpful for numerical solutions of
flow equations. In the generic case, one two-parameter function has to
be fitted in order to describe the flows \eq{ThresholdT} or
\eq{ThresholdTF}. In turn, only two one-parameter functions are needed
once they factorise as in \Es{ldnT} and \eq{ldnTF}. This
simplification is substantial, and even more so, because the functions
\eq{ldnT} and \eq{ldnTF} have a very simple {analytical} form.\step

%********|*********|*********|*********|*********|*********|*********|****
\section{Derivative expansion to second order}\label{Anomalous}
%********|*********|*********|*********|*********|*********|*********|****

In this section, we apply the optimised regulator to higher order in
the derivative expansion. We first discuss the general structure of
the equations. Furthermore, we show that a simple factorisation of the
flow takes place for field-independent wave function renormalisation. The
physical origin of the factorisation is discussed, and its realisation
on the level of the flow equation leads, as in the thermal case, to
better convergence properties of the flow and the derivative
expansion. For technical details on the computations, we defer to the
Appendices .

%********|*********|*********|*********|*********|*********|*********|****
\subsection{Wave-function renormalisation}
%********|*********|*********|*********|*********|*********|*********|****

In the preceding sections we have restricted the discussion to the
leading order in a derivative expansion. This implied the vanishing of
the anomalous dimensions $\eta_\phi=\eta_\psi=0$ or $Z_\phi\equiv
Z_\psi\equiv 1$.  To higher order in the derivative expansion the
multiplicative renormalisation of the fields has to be taken into
account, according to
\bea
\label{Zphi} \phi &\to& Z^{1/2}_{\phi,k}\phi\\
\label{Zpsi} \psi &\to& Z^{1/2}_{\psi,k}\psi\,.
\eea
Once higher derivative terms are included in the Ansatz for the
effective action, additional flow equations for the corresponding
coefficient functions like $Z_{\phi,k}(q^2,\bar\rho)$ and
$Z_{\psi,k}(q^2,\bar\rho)$ have to be studied. The wave function
renormalisations are functions of the scale parameter $k$ and can
depend as well on momenta $q^2$ or on the mean fields $\bar \rho$. To
second order in the derivative expansion the wave function
renormalisation is evaluated at a particular momentum scale
$q^2=k_0^2$ which fixes the renormalisation conditions. Typical
choices for $k_0$ are $k_0=0$ and $k_0=k$. \step

The most important new ingredient at this order is the scale- and
field-dependence of the wave function renormalisations.  In the
example defined through \Eq{AnsatzGamma}, these are given by the
functions $Z_k$ and $Y_k$.  Here, the function $Z_k$ is responsible
for the renormalisation of the $N-1$ ``Goldstone'' modes, which differs
from $\tilde Z_k=Z_k+\bar \rho Y_k$ for the ``radial'' mode. The fact
that different wave function renormalisations appear to second order
in the derivative expansion (depending of the theory considered) is of
no relevance for the following discussion of the flows. The parametric
dependence of the flow on either $Z_k$ or $\tilde Z_k$ is the same.
\step

Let us introduce an additional function $z_k(\rho)$ as
\beq\label{z}
Z_k(\bar\rho)=Z_k(\bar\rho_0)\, z_k(\bar\rho)\,.
\eeq 
We have factored out a constant term $Z_k(\bar\rho_0)$ chosen at an
arbitrary reference point. We have chosen the reference point
$\bar\rho=\bar\rho_0$ which fixes the renormalisation of the fields
for all momenta. Typical choices are either $\rho_0=0$, or $\rho_0=$
the minimum of the scale-dependent potential. The split \eq{z} allows
to separate the non-trivial field-dependence, contained in
$z_k(\bar\rho)$, from an overall renormalisation contained in
$Z_k(\bar\rho_0)$. The factor $z_k$ is normalised as
$z_k(\bar\rho_0)=1$. \step

In order to provide a simple form for the flow it is useful to
introduce the field-independent factor $Z_k(\bar\rho_0)$
into the regulator function,
\bea\label{R_Z}
R_k    &\to& Z_{\phi,k}R_k\\
R_{F,k}&\to& Z_{\psi,k}R_{F,k}\,.
\eea 
The flow equation, when written in terms of renormalised variables
\eq{Zphi} and \eq{Zpsi}, receives additional contributions
proportional to the anomalous dimensions
\bea\label{AnomalousDimB}
\eta_\phi&=&-\partial_t \ln Z_{\phi,k}\,, \\
\label{AnomalousDimF}
\eta_\psi&=&-\partial_t \ln Z_{\psi,k} \,, 
\eea
because the derivative $\partial_t R_k$ in the flow equation now acts
also on the explicit scale-dependence contained in $Z_k$.

%********|*********|*********|*********|*********|*********|*********|****
\subsection{Generic flows to second order}
%********|*********|*********|*********|*********|*********|*********|****

To second order in the derivative expansion the flow has turned into a
function of the field amplitudes $\omega$, the anomalous dimension
$\eta$ and the field dependent function $z$. The corresponding bosonic
flow is defined as
\beq\label{Thresholdz}
\ell^d_0(\omega,z,\eta)=\int^\infty_0 dy y^{\s0d2-1}
\0{-y^2\,r'(y)-\s012 \eta\, y\, r(y)}{y[z+r(y)]+\omega}\,.
\eeq
Notice that the pole structure of the flow is changed, owing to the
function $z(\rho)$. The effective inverse propagator becomes a
function of the fields:
\beq\label{Polez}
P^2(\rho,y)=y[z(\rho)+r(y)]\,.
\eeq
The location of the pole of \Eq{Thresholdz} at $-\omega =
C(\rho)=\min_{y\ge 0}P^2(\rho,y)$ has turned into a function of the
fields. For the optimised regulator \eq{ropt} the pole is located at
$C=\min\{1,z\}$.  Compared to the leading order in the derivative
expansion, the pole structure is modified once $z\neq 1$. \step

For the fermionic case, the flows are given as
\beq\label{ThresholdzF}
\ell^d_{F,0}(\omega,z_F,\eta_\psi)=\int^\infty_0 dy y^{d/2}
\0{-(2y\,r_F'+\eta_\psi\,r_F) (z_F+r_F)}{y[z_F+r_F(y)]^2+\omega}\,.
\eeq
The pole structure changed as well, as follows from
\beq\label{PolezF}
P_F^2(\rho,y)=y[z_F(\rho)+r_F(y)]^2\,.
\eeq
The location of the pole of \Eq{Thresholdz} at
$-\omega=C_F(\rho)=\min_{y\ge 0}P_F^2(\rho,y)$ has turned into a
function of the fields. For the regulator \eq{roptF} the pole is
located at $C_F=\min\{1,z_F\}$. The pole structure is modified
compared to the leading order in the derivative expansion once
$z_F\neq 1$. \step

%********|*********|*********|*********|*********|*********|*********|****
\subsection{Optimised flows to second order}\label{Opt-2nd}
%********|*********|*********|*********|*********|*********|*********|****

We now turn to the optimised flows and discuss their structure at
second order in the derivative expansion. We refer to the Appendices
for all technical details.\step

In the case of a generic wave-function renormalisation where
$z_k(\bar\rho)$ is a non-trivial function of the fields, the function
$\ell^d_0(\omega,z,\eta)$ as defined in \Eq{Thresholdz} can be
evaluated explicitly for the regulator \eq{Ropt}. The structure of the
flow is as follows. Consider the denominator of \Eq{Thresholdz}, given
by $y(z+r)+\omega$. It can be rewritten as
$1+\omega+y(z-1)+(y(1+r)-1)$. The last term in brackets vanishes for
the optimised regulator \eq{Ropt} because the integration is
restricted to $y\le 1$. The remaining term can be written as the
product $(1+\omega)(1-\s0{1-z}{1+\omega}y)$. Notice also that the
numerator of \Eq{Thresholdz} neither depends on $z$ nor on $\omega$.
These observations lead to the following conclusions. First, and apart
from an overall $\omega$-dependence $\sim (1+\omega)^{-1}$, the
optimised flow depends on $z$ and $\omega$ only through the variable
\beq\label{xi}
\xi\equiv\0{1-z}{1+\omega}\,.
\eeq 
Second, the optimised flow factorises into a leading order 
term \eq{ld0} and a remaining factor $B_d(\xi,\eta)$,
\beq\label{Eta-Correction}
\ell^d_0(\omega,\xi,\eta)=
\ell^{d}_0(\omega)\,
B_d\left(\xi,\eta\right)\,,
\eeq
because the denominator of \Eq{Thresholdz} contains a
momentum-independent factor $(1+\omega)$. Here we have introduced
$\ell^d_0(\omega,\xi(\omega,z),\eta)\equiv\ell^d_0(\omega,z,\eta)$.
It it interesting to see that the structure of the optimised flow is
still quite simple. An integral representation of $B_d(\xi,\eta)$ is
given in \Eq{Bd-integral}. (For all $\xi<1$, $B_d(\xi,\eta)$ can be
expressed in terms of hypergeometric functions, cf.~\Es{l-wze},
\eq{l-wze-1} and \eq{l-wze-2}; closed expressions of
\Eq{Eta-Correction} for $d=4,3$ and $2$ dimensions are given in
\Es{l40res}, \eq{l30res} and \eq{l20res}, respectively.) For $|\xi|<1$
the function $B_d(\xi,\eta)$ can be Taylor-expanded in arbitrary
dimensions, to wit
\beq\label{ld0Series}
B_d\left(\xi,\eta\right)
=
\sum^\infty_{n=0}
\0{\xi^n}{1+\s0{2n}{d}}
\left(1-\0{\eta}{d+2+2n}\right)
=
1-\0{\eta}{d+2}+\0{\xi}{1+\s02d}+{\cal O}(\xi^2,\xi\eta)\,.
\eeq
The
series representation \eq{ld0Series} is best suited for the flow as
long as $|\xi|$ remains small. This corresponds to either the limit of a
field-independent wave function renormalisation $z(\rho)\equiv 1$, or, for
any $z$, to the limit of large amplitudes $\omega$. From
\Eq{ld0Series}, we obtain for \Eq{Eta-Correction} to zeroth order in
$\xi$
\beq\label{ld0Eta}
\ell^d_0(\omega,\xi,\eta)=
\ell^d_0(\omega)\left(1-\0{\eta}{d+2}\right)
+{\cal O}\left(\0{\xi}{1+\omega},\0{\xi\eta}{1+\omega}\right)\,.
\eeq
We note that the $\eta$-dependent correction in \Eq{ld0Eta} has the
{same} functional dependence on the amplitude $\omega$ as
\Eq{ld0}. Stated differently, the optimised regulator leads to a
simple factorisation in both the decoupling limit $\omega\gg 1$ and
for the case of a field-independent anomalous dimension $\xi\equiv 0$.\step

For completeness we give also the result for the bosonic flow at
finite temperature. The corresponding flow
$\ell^d_0(\omega,\xi,\tau,\eta)$ still factorises as
\beq\label{ld0EtaXiT}
\ell^d_0(\omega,\xi,\tau,\eta)=B_d(\xi,\tau, \eta)\ell^d_0(\omega)\,.
\eeq
This is the generalisation of \Eq{ldnT} to second order in the
derivative expansion. We only have to replace the function
$B_d(\xi,\eta)$ by its temperature-dependent counterpart
$B_d(\xi,\tau, \eta)$ [cf.~\Eq{Bd-general}]. It is
straightforward, if tedious, to establish explicitly that the function
$B_d(\xi,\tau, \eta)$ represents dimensional reduction in precisely
the same way as $B_d(\tau)\equiv B_d(\xi=0,\tau, \eta=0)$. Let us
consider the most interesting case, which is the leading order in
$\xi\ll 1$. In this limit, \Eq{ld0EtaXiT} reads
\beq\label{ld0EtaT}
B_d(\xi,\tau,\eta)
=
B_d(\tau, \eta)
+{\cal O}\left(\xi,\xi\eta\right)\,.
\eeq
The function $B_d\left(\tau, \eta\right)$ can be expressed as
\beq\label{Bd-eta}
B_d\left(\tau, \eta\right)=
B_d\left(\tau\right)
-\eta\ 
\0{d}{d^2-1}
\0{v_{d-1}}{v_d}  
\0{\tau}{2\pi}
\sum_m \left(1-c_m^2\tau^2\right)^{\s0{d+1}{2}}
\Theta\left(1-c_m^2\tau^2\right)\,.
\eeq
The function $B_d(\tau)$, \Eq{Bd}, has been discussed in
Section~\ref{Thermal}. The new ingredient, beyond leading order, are
given by the corrections $\sim\eta$ in \Eq{Bd-eta}. Every single
Matsubara mode contributes as
\beq\label{singleEta}
\sim \tau \left(1-c_m^2\tau^2\right)^{\s0{d+1}{2}}
\Theta\left(1-c_m^2\tau^2\right)\,.
\eeq
Compared to the leading order contributions \eq{single}, we notice
that \Eq{singleEta} follows from \Eq{single} for $d\to d+2$. The
reason is very simple. In the flow equation, the anomalous dimension
is proportional to a term containing an additional factor $\sim q^2$,
which effectively increases the momentum measure by two dimensions.
This has an immediate consequence. The thermal decoupling in
\Eq{Bd-eta} proportional to $\sim \eta$ is much smoother than the
leading order decoupling, simply because the spikes are less
pronounced the higher the dimension. Therefore, the spikes observed in
Fig.~2 are smoothed-out once $\eta$ (and $\xi$) are non-vanishing.
\step
 
In the opposite regime where $\s0{|1-z|}{1+w}\gg 1$, only a few leading
terms of the series \eq{ld0Series} have to be retained. This limit is
of relevance close to the pole region of the flows $\omega\to -1$, or 
in the region
of large $z\gg 1$. From the explicitly resummed expressions \eq{l40res},
\eq{l30res} and \eq{l20res}, we conclude that a factorisation as
\beq\label{ld0Large}
\ell^d_0(\omega,z,\eta)=f_d(\omega,z)\left(1-\0{\eta}{d}\right)\,
\eeq
holds true, and $f_4(z)=(z-1)^{-1}$, $f_3(z)=2f_4(z)$ 
and $f_2(\omega,z)=f_4(z)\ln(\s0{z+\omega}{1+\omega})$. \step

It is not surprising that a similar structure is found for fermionic
flows.  The correction term due to the substitution \Eq{R_Z}
simplifies \Eq{ThresholdzF} to
\beq\label{EtaF-Correction}
\ell^d_{F,0}(\omega,z_F,\eta_\psi)
=\int_0^1 dy y^{d/2-1}\,
\0{[1+\sqrt{y}(z_F-1)][1-\eta_\psi(1-\sqrt{y})]}{[1+\sqrt{y}(z_F-1)]^2+w}\,.
\eeq
\Eq{EtaF-Correction} factorises as 
\beq\label{EtaF-factorisation}
\ell^d_{F,0}(\omega,z_F,\eta_\psi)
=\ell^d_{F,0}(\omega)\, F_d(\omega,z_F,\eta_\psi)\,.
\eeq
The function $F_d(\omega,z,\eta)$ can be expressed in terms of
hypergeometric functions. At finite temperature, and for $z=1$ and
$\eta=0$, it reduces to \Eq{Fd}. Here, we are only interested in the
structure of the flow for a nearly field-independent wave function
renormalisation, $z\approx 1$, or for the decoupling limit. We find
\beq\label{ld0EtaF}
F_d(\omega,z_F,\eta_\psi)=
1-\0{\eta_\psi}{d+1}
 +\0{d(z_F-1)}{d+1}\left(1-\0{2}{1+\omega}\right)
  \left(1-\0{\eta_\psi}{d+2}\right)
 +\cdots 
\eeq
The two leading terms in \Eq{ld0EtaF} show that also fermionic flows
factorise for field-independent wave function renormalisation.\step

%********|*********|*********|*********|*********|*********|*********|****
\subsection{Discussion}
%********|*********|*********|*********|*********|*********|*********|****

The structure of the flow has increased to second order in the
derivative expansion. Let us discuss first the case of a
field-independent wave function renormalisation $z\equiv 1$. The
corresponding flows \eq{ld0Eta} and \eq{ld0EtaF} for the optimised
regulators factorise, similar to the thermal case to leading order in
the derivative expansion.  Physically speaking, this structure can be
made plausible as follows.  The flow, when written in terms of the
{renormalised} fields -- and under the assumption that the
renormalisation is momentum- and field-independent -- depends, in
addition to the fields, only on the anomalous dimension. The anomalous
dimension is field independent, and, in consequence, unable to
distinguish between fields of different amplitudes contained in
$\omega$, which parametrise the quantum fluctuations. Therefore, it is
natural that the flow factorises the contributions induced through
$\eta$ from those induced by the amplitudes $\omega$. The
disentanglement is realised by the optimised regulators.\footnote{From
the definition of \Es{Thresholdz} and \eq{ThresholdzF} it follows
that {all} homogeneous regulators with $r(y)\sim y r'(y)$ (or
$r_F(y)\sim y r_F'(y)$, respectively) factorise the anomalous
dimension from the field-dependent part of the flow.}\step

In turn, a generic flow does not reflect this factorisation. Rather,
it leads to an entanglement between the renormalisation of the
effective potential induced by the infrared regulator, and the
renormalisation parametrised by a field-independent anomalous dimension.
This is immediately evident from the observation that the
$\eta$-dependent and the $\eta$-independent contributions to the flow
of the effective potential have {different} functional forms as
functions of the fields. At this level, the entanglement is due to the
regulator, which modifies the coupling amongst all operators of the
effective action. As mentioned in the thermal case, the entanglement
is of no importance for the {full} solution to the flow. In turn,
the factorisation is very useful for {approximate} solutions. It
leads to more stable flows because irrelevant couplings, entirely due
to the regulator, are removed. The same reasoning as given at the end
of the previous section applies.\step

For the thermal bosonic flow \eq{ld0EtaXiT}, we notice that the
dependence on the anomalous dimension enters the thermal factor
$B_d(\tau,\eta)$. In particular, the thermal corrections do not
factorise from those due to a field-independent anomalous dimension. This
structure can be understood as follows. The wave function
renormalisation enters the momentum trace as a multiplicative
renormalisation proportional to the kinetic term $q^2$. At finite
temperature within the imaginary time formalism, the spatial and the
temporal loop momenta are treated in an unequal way. Hence, thermal
fluctuations couple in a non-trivial manner to the anomalous dimension
of the fields. This implies that the temperature-dependent factor
itself is modified due to the anomalous dimension, which provides the
physical reason way no factorisation of the temperature effects from
the anomalous dimension are expected in the first place. \step

For the case of a field-dependent wave-function renormalisation, a
simple factorisation similar to \Eq{ld0Eta} is not expected, simply
because the wave function renormalisation is a function of the fields.
Hence, the wave function renormalisation can distinguish different
field amplitudes, in contrast to the field-independent case. However, two
observations are still worth mentioning. First of all, we observe a
{\it partial factorisation}, which is evident from \Es{Eta-Correction}
and \eq{ld0EtaXiT}. This structure is based on the fact that the
$z$-dependence enters only through the variable \eq{xi}, as opposed to
the generic case. For $\s0{|1-z|}{1+w}\ll 1$, only a few leading terms
have to be retained from the explicit series \eq{ld0Series}. It
follows that each power of $\s0{|1-z|}{1+w}$ is renormalised
proportional to the anomalous dimension and an order-dependent
numerical coefficient. Secondly, the limit for $\s0{|1-z|}{1+w}\gg 1$
again allows for a simple factorisation, as follows from
\Eq{ld0Large}. Here, the wave function renormalisation can no longer
distinguish field amplitudes, allowing for this simple structure.\step

A final comment concerns the numerical prefactors $\sim\eta$ as found
in \Es{ld0Eta} and \eq{ld0EtaF}. We emphasise that the coupling of the
anomalous dimensions to the effective potential is, apart from the
field dependence, {\it dimensionally} suppressed -- by factors
$\s01{d+2}$ for bosons and $\s01{d+1}$ for fermions -- as opposed to
the leading order contributions. This additional suppression is
noteworthy because the convergence of the derivative expansion is
controlled by small anomalous dimensions of the fields. Here, we have
just shown that an expansion performed with an optimised regulator
leads to an {\it additional} dimensional suppression of the
back-coupling of the anomalous dimension to the effective potential. A
more detailed discussion of this observation will be given elsewhere.

%********|*********|*********|*********|*********|*********|*********|****
\section{ Proper-time regularisation}\label{HeatKernel}
%********|*********|*********|*********|*********|*********|*********|****

In this section we leave aside the conceptual framework of the ERG
based on a momentum-scale regularisation and address flows based on an
operator cut-off regularisation. Our aim is to provide the analog of
the optimised regulator \eq{Ropt} within the proper-time
regularisation method. For a more detailed comparison with the exact
renormalisation group, we refer the reader to
Ref.~\cite{Litim:2001hk}.\step

A simple flow has been derived from a one-loop expression for the
effective action which is UV and IR regularised using a Schwinger
proper-time representation of the operator trace
\cite{Schwinger:1951nm}, amended by a regulator function
$f^{(d)}_k(\Lambda,s)$ within the proper-time integral
\cite{Oleszczuk:1994st}. The flow with respect to the infrared scale
parameter $k$ follows from a 1-loop improvement as \cite{Liao:1996fp}
\beq\label{ProperFlow}
\partial_t \Gamma_k=
-\s012 \int_0^\infty \0{ds}{s}
\left( \partial_t f^{(d)}_k(\Lambda,s)\right) 
\Tr\exp\left(-s\Gamma^{(2)}_k\right)\,.
\eeq
We refer to this flow as the ``proper-time renormalisation group''
(PTRG). It describes the partial resummation of perturbative diagrams.
The proper-time regulator function plays the role of the momentum
regulator $R_k$ within the ERG. The flow \eq{ProperFlow} is governed
by the IR scale $k$. Following Ref.~\cite{Liao:1996fp}, we introduce a
dimensionless function $f(x)$ as $f_k^{(d)}(\Lambda,s)=f(\Lambda^2
s)-f(k^2 s)$ and require $f(x\to\infty)=1$ and $f(x\to 0)=0$. This
ensures that the usual Schwinger proper time representation is reached
in the UV limit.\step

We are not aware of a simple and generic optimisation criterion,
analogous to \Eq{Opt}, which derives from within the PTRG formalism.
Furthermore, the flow \eq{ProperFlow} has no path integral derivation,
which makes a conceptual reasoning much more difficult. However, it is
still possible to show that a function $f_{\rm opt}(x)$ exists which
is equivalent to the optimised ERG regulator \eq{Ropt} to the leading
order in the derivative expansion.\step

To that end, we apply \Eq{ProperFlow} to a $N$-component real scalar
theory in $d$ dimensions and to leading order in the derivative
expansion. Using the Ansatz \eq{AnsatzGamma} the flow for the
effective potential $U_k(\bar\rho)$ with $\bar\rho=\s012\phi_a\phi_a$
becomes
\beq\label{ProperUflow}
\partial_t U_k(\bar\rho)=\s012(4\pi)^{-\s0d2}
\int_0^\infty\0{ds}{s^{1+d/2}}
\partial_t f^{(d)}_k(\Lambda,s)
\left[
      e^{-s\left(U'_k(\bar\rho)+2\bar\rho U''_k(\bar\rho)\right)}
+(N-1)e^{-sU'_k(\bar\rho)}
\right]\,.
\eeq
This flow is {identical} in form to the ERG flow
\eq{FlowUThreshold}, if we replace the ERG flow in \Eq{FlowUThreshold}
by the proper-time flow
\beq\label{ProperThreshold}
\ell^d_{0}(\omega)=\s012 \Gamma(\s0d2)\int_0^\infty
dx x^{-1-\s0d2} 
\large( \partial_t f(x)\large) \exp (-x \omega)\,.
\eeq
Here, the integration variable is $x=k^2s$ and stems from the
proper-time integration, in contrast to \Eq{ThresholdDef}, where
$y=q^2/k^2$ stems from the momentum trace. Now, consider a specific
class of proper-time regulator functions:
\beq\label{Ansatz}
f(x)=\0{\Gamma(m,x)}{\Gamma(m)}\,,\quad\quad
\partial_t f(x)=
\0{2x^m e^{-x}}{\Gamma(m)} 
\eeq
We have introduced a free parameter $m$ describing different
regulators, and the incomplete $\Gamma$-function $\Gamma(m,x)=\int^x_0
dy y^{m-1} e^{-y}$. This yields the simple expression
\beq\label{PT-Threshold}
\ell^d_{0}(\omega)=
\0{\Gamma(m-\s0d2)\Gamma(\s0d2)}{\Gamma(m)}
(1+\omega)^{\s0d2-m}
\eeq
which agrees with \Eq{ld0} for $m=1+\s0d2$, or
\beq\label{ProperOpt}
f^{\rm opt}(x)=\0{\Gamma(\s0d2+1,x)}{\Gamma(\s0d2+1)}\,,\quad\quad
\partial_t f^{\rm opt}(x)=
\0{2x^{1+d/2} e^{-x}}{\Gamma(\s0d2+1)}\,. 
\eeq
The optimised proper-time regulator \eq{ProperOpt} corresponds to the
optimised regulator \eq{Ropt} within the ERG approach. Hence, it is
possible to identify an optimal regulator function for proper-time
flows, owing to their close similarity to the ERG to leading order in
the derivative expansion.

%********|*********|*********|*********|*********|*********|*********|****
\section{Conclusions and outlook}\label{Discussion}
%********|*********|*********|*********|*********|*********|*********|****

This study was motivated by two observations. First, an application of
the ERG to realistic physical problems is bound to certain
approximations. Second, approximate solutions of flow equations depend
spuriously on the infrared regulator. Combining these observations, it
became obvious that an understanding of the spurious scheme dependence
is mandatory in order to provide predictive power for approximate
solutions. Previously, we showed that the gap of the full inverse
propagator controls convergence properties of approximate solutions
\cite{Litim:2000ci}. It has also been shown, based on the computation
of critical exponents for the Ising universality class, that the
convergence of the derivative expansion is controlled by the gap
\cite{Litim:2001fd}. These observations lead to the conclusion that the
freedom in the choice for the IR regulator can be used to maximise the
physical information contained within a given approximation or
truncation. \step

An interpretation of the interplay between the RS function on one
side, and convergence of approximate flows on the other, is as
follows. The IR regulator -- by regulating the flow -- modifies the
interactions at intermediate scales $k\neq 0$ amongst all operators of
the theory. Eventually, these cancel out for the integrated full flow,
but not for approximated ones.  Hence, changing the RS function for an
approximated flow modifies some remaining RS dependent terms which
cannot be cancelled due to the missing contributions from neglected
operators. Therefore, a ``fine-tuning'' of the RS function allows to
partly incorporate higher-order effects within the lower orders of a
given approximation.  This corresponds to an optimisation. \step

The present derivation of optimised ERG flows had two ingredients.
First, we made use of a generic optimisation criterion for bosonic and
fermionic fields \cite{Litim:2000ci}, which states that the gap of the
full inverse propagator, as a function of momenta, should be as large
as possible. This way, the ERG flow is the least singular, and
approximations to such flows are expected to be much more stable,
leading to improved convergence of expansions. Second, we added the
specific requirement that the effective inverse propagator be
momentum-independent in the IR region (cf.~Fig.~1). \step

We have studied specific optimised ERG flows for bosonic or fermionic
theories up to second order in the derivative expansion and at
vanishing and non-vanishing temperature. Their specific properties
have been discussed in detail at the end of the corresponding
sections. Here, let us only mention the perhaps most surprising
property of optimised flows, which is the disentanglement of thermal
and quantum fluctuations to leading order in the derivative expansion.
A similar factorisation occurs for field-independent wave function
renormalisations. \step

More generally, optimised ERG flows owe their main properties to the
``flatness'' of the effective inverse propagator, which extends over
the entire momentum region $q^2\le k^2$ for the specific regulator
studied here.  Other regulators can lead to similar factorisation and
convergence properties. Prime candidates are given by solutions to the
optimisation criterion: as is evident from Fig.~1, they automatically
lead to flat effective propagators -- at least within a small region
about the minimum of the effective inverse propagator. If this region
extends over the domain where the flow equation receives its main
contributions, we expect to find equally good flows.  \step

An important conclusion is that the optimisation ideas discussed here
should be useful for high-precision computations based on this
formalism.  Increasing the precision normally implies a full
computation at the following order of the expansion. Here, we argued
that the physical results can be improved already within a fixed order
of the expansion. Immediate applications of optimised flows concern
the computation of universal critical exponents for $N$-component
scalar theories in three dimensions, or the study of convex effective
potentials within a phase with spontaneous symmetry breaking. On the
conceptual side, it is possible to show that the optimisation
criterion can be interpreted as a natural minimum sensitivity
condition, somewhat similar to the principle of minimum sensitivity as
employed within perturbative QCD.  We will leave a detailed discussion
of these results to a future
 publication \cite{Litim:2001fd}.  \step

Our analysis can be extended in a number of directions. For gauge
theories, modified Ward or BRST identities ensure the gauge invariance
of physical Green functions \cite{gaugefields}, and the optimisation
criterion is compatible with such additional constraints. This
optimisation can also be implemented for field theories at finite
temperature within the real-time formalism \cite{Litim98}. While our
present analysis is based on the derivative expansion, it seems
worthwhile to study optimisations for other systematic expansions like
expansions in powers of the fields. Finally, it would be interesting
to see how these ideas apply to flows at finite density or to
Hamiltonian flows \cite{Wegner:1994}.\\[4ex]

\noindent
{\bf Acknowledgement:} 
This work has been supported by a Marie-Curie fellowship under EC
contract no.~HPMF-CT-1999-00404.\\[4ex]

%********|*********|*********|*********|*********|*********|*********|****
\setcounter{section}{0}
\renewcommand{\thesection}{Appendix \Alph{section}}
\renewcommand{\theequation}{\Alph{section}.\arabic{equation}}
%********|*********|*********|*********|*********|*********|*********|****

%********|*********|*********|*********|*********|*********|*********|****
\section{Flows to second order in the derivative expansion}
%********|*********|*********|*********|*********|*********|*********|****

In this Appendix, we derive explicit expressions for the optimised
flow for the effective potentials to second order in the derivative
expansion at both vanishing and non-vanishing temperature.\step

Our starting point is the flow for the effective potential to second
order in the derivative expansion. We consider the bosonic flow
at finite temperature. All information about the flow is parametrised by
\beq\label{l-start}
\ell^d_0(\omega,z,\tau,\eta)=
\0{v_{d-1}}{v_d}\0{\tau}{2\pi}\sum_n 
\int_0^{1-c^2_n\tau^2} dy y^{\s0{d-3}{2}}
\0{[1-\s012\,\eta\,(1-y-c_n^2\tau^2)]\,\Theta(1-c_n^2\tau^2)}{(z-1)(y+c_n^2\tau^2)+1+\omega}
\eeq
Here, $\omega(\rho)$ is the field variable, $z(\rho)$ a field
dependent wave function renormalisation, $\eta$ the anomalous
dimension, $\tau=2\pi T/k$ the rescaled dimensionless temperature,
and $c_n=n$ the Matsubara modes in the
bosonic case, \Eq{cmb}. The constants $v_d$ are defined in
\Eq{vd}.\step

The leading order behaviour of \Eq{l-start} is given by the function
\beq\label{l-basic}
\ell^d_0(\omega)=\02d\01{1+\omega}\,,
\eeq
which follows from \Eq{l-start} for $z=1$, $\tau=0$ and $\eta=0$. 
Factorizing the main building block \Eq{l-basic} from \Eq{l-start}, 
we notice that the remaining factor depends on both $\omega$ and the 
variable $z$ only through the combination
\beq\label{trafo}
\xi\equiv \0{1-z}{1+\omega}\,.
\eeq
Therefore, it is most natural to make the variable transform 
\Eq{trafo} by writing
$\ell^d_0(\omega,z,\tau,\eta)\equiv
\ell^d_0(\omega,\xi(z,\omega),\tau,\eta)$, and to rewrite the flow
\eq{l-start} as
\beq\label{l-factorisation}
\ell^d_0(\omega,\xi,\tau,\eta)=
\ell^{d}_0(\omega)\,
B_d\left(\xi,\tau,\eta\right)\,,
\eeq
where
\beq\label{Bd-general}
B_d(\xi,\tau,\eta)=
\0{d}{2}
\0{v_{d-1}}{v_d}\0{\tau}{2\pi}\sum_n 
\int_0^{1-c^2_n\tau^2} \!dy\, y^{\s0d2-\s032}\,
\0{1-\s0{\eta}{2}\,(1-y-c_n^2\tau^2)}{1-\xi\,(y+c_n^2\tau^2)}
\,\Theta(1-c_n^2\tau^2)
\eeq
Below, if not stated otherwise, we adopt a simplified notation:
functions are evaluated at the points $\xi=0$, $z=1$, $\tau=0$ or
$\eta=0$, if the corresponding arguments are not displayed. With these
definitions at hand, we can face the explicit computation of
\Eq{l-start}. \step

Let us compute the function $B_d(\xi,\tau,\eta)$ more explicitly. 
Since the anomalous
dimension enters only linear in \Eq{l-start}, it is helpful to rewrite
\Eq{Bd-general} as
\beq\label{b-factorisation}
B_d\left(\xi,\tau,\eta\right)=
B_d\left(\xi,\tau\right)
-\eta\bar B_d\left(\xi,\tau\right)\,.
\eeq
For $\xi<1$, the remaining integration over the momentum variable 
in \Eq{Bd-general} can be performed. This leads to
\bea
B_d\left(\xi,\tau\right)
&=&
\0{d}{d-1}
\0{v_{d-1}}{v_d}
\0{\tau}{2\pi}
\sum_n  A_n(\xi,\tau)
(1-c_n^2\tau^2)^{\s0{d-3}{2}}\Theta(1-c_n^2\tau^2)\times 
\nonumber \\ \label{b-rt}  && 
\,{}_2F_1\left(1,\s0{d-1}{2};\s0{d+1}{2};\xi A_n(\xi,\tau)\right)
\eea
The function 
\beq\label{2F1-Def}
{}_2F_1(a,b;c;z)=
\0{\Gamma(c)}{\Gamma(b)\Gamma(c-b)}
\int^1_0 dt\ t^{b-1}(1-t)^{c-b-1}(1-tz)^{-a}
\eeq
with ${}_2F_1(a,b;c;z)={}_2F_1(b,a;c;z)$ denotes the hypergeometric 
function \cite{2F1}. We also introduced the thermal amplitude function
\beq
\label{Am}
A_n(\xi,\tau)=\0{1-c_n^2\tau^2}{1-\xi \, c_n^2\tau^2}\,.
\eeq
The index $n$ corresponds to the Matsubara mode. The factors $A_n$
only appear in combination with the factor $\Theta(1-c_n^2\tau^2)$. At
the limits,
\bea
\label{At0}
A_n(\xi,0)=A_0(\xi,\tau)&=&1\\
\label{At1}
A_n(\xi,c_n^{-1})&=&0\,.
\eea
In the same way, we find from \Eq{Bd-general} for the term $\sim\eta$ in
\Eq{l-factorisation} the explicit expression
\bea
\bar B_d\left(\xi,\tau\right)&=&
\012
\0{d}{d-1}
\0{v_{d-1}}{v_d}
\0{\tau}{2\pi}
\sum_n A_n(\xi,\tau)(1-c_n^2\tau^2)^{\s0{d-1}{2}}\Theta(1-c_n^2\tau^2)
\times \nonumber \\ && \label{bb-rt}
\left[
{}_2F_1\left(1,\s0{d-1}{2};\s0{d+1}{2};\xi A_n(\xi,\tau)\right)
- 
\s0{d-1}{d+1}\
{}_2F_1\left(1,\s0{d+1}{2};\s0{d+3}{2};\xi A_n(\xi,\tau)\right)
\right]
\eea
Combining \Eq{b-rt} with \Eq{bb-rt} gives \Eq{b-factorisation} and
hence \Eq{l-factorisation} explicitly. \Eq{l-factorisation} is the
most general expression for the factorisation at second order in the
derivative expansion and at finite temperature.\step

The temperature dependence of the function
$B_d\left(\xi,\tau,\eta\right)$ describes dimensional reduction. In
particular, it obeys the limits
\bea\label{Bd-lowT}
B_d\left(\xi,\tau=0,\eta\right)&\equiv &
B_d\left(\xi,\eta\right)\\
\label{Bd-highT}
B_d\left(\xi,\tau\ge 1,\eta\right)&=&
\0{v_{d-1}}{v_d}\0{d}{d-1}\0{\tau}{2\pi}
B_{d-1}\left(\xi,\eta\right)\eea
The low- and high-temperature limits \eq{Bd-lowT} and \eq{Bd-highT}
are discussed in the following appendices.\step

Let us consider the case where $\xi=0$. It corresponds either to the
case of a field-independent wave function renormalisation $z\equiv 1$,
and/or to the decoupling regime $\omega\gg 1$. A few properties of
\Eq{l-factorisation} have been discussed in the main text. For $\xi=0$
the factor \eq{b-factorisation} reduces to
\beq\label{Bdeta}
B_d\left(\tau, \eta\right)=
\0{d}{d-1} 
\0{v_{d-1}}{v_d}  
\0{\tau}{2\pi}
\sum_n \left(1-c_n^2\tau^2\right)^{\s0{d-1}{2}}
\left( 1 - \eta\0{1-c_n^2\tau^2}{d+1}\right)
\Theta\left(1-c_n^2\tau^2\right)\,.
\eeq
This corresponds to \Eq{Bd} discussed in Section~\ref{Thermal}. In
addition, we notice that
\bea
\label{Bdeta0}
B_d(\tau,\eta\to 0)&=& B_d(\tau)\\
\label{Bdtau0}
B_d(\tau\to 0, \eta)&=&1 - \0{\eta}{d+2}\\
\label{Bdtau1}
B_d(\tau\ge 1, \eta)&=&\0{d}{d-1} \0{v_{d-1}}{v_d}  \0{\tau}{2\pi}
\left( 1 - \0{\eta}{d+1}\right)\,.
\eea
\Eq{Bdeta0} corresponds to \Eq{ld0}, \Eq{Bdtau0} to the
low-temperature limit \eq{ld0Eta}, and \Eq{Bdtau1} to the
high-temperature limit \eq{Reduction1}. \step

Similar results are found for the fermionic case, though not discussed
explicitly.

%********|*********|*********|*********|*********|*********|*********|****
\section{Low temperature limit}\label{Low}
%********|*********|*********|*********|*********|*********|*********|****

In the low temperature limit $\tau\to 0$, the flow \eq{l-start}
simplifies to
\beq\label{l-low}
\ell^d_0(\omega,z,\eta)=
\int_0^{1} dy y^{\s0{d-2}{2}}
\0{1-\s012\,\eta\,(1-y)}{(z-1)y+1+\omega}\,.
\eeq
The remaining integration in \Eq{l-low} is solved to give
\beq\label{l-low-Series}
\ell^d_0(\omega,\xi,\eta)=
\ell^d_0(\omega)\,B_d(\xi,\eta)
\eeq
with
\beq\label{Bd-integral}
\,B_d(\xi,\eta)=
\0d2\int^1_0 dy\,y^{\s0d2-1}
\left(\0{1}{1-\xi y}-\0{\eta}{2}\0{1-y}{1-\xi y}\right)
\eeq
in arbitrary dimensions. For $\xi\equiv\s0{1-z}{1+\omega}< 1$, and 
hence $z>-\omega$, the integration can be peformed analytically. We find
\beq\label{l-wze}
B_d(\xi,\eta)=
B_d(\xi)-\eta\,\bar B_d(\xi)
\eeq
where
\beq\label{l-wze-1}
B_d(\xi)=
{}_2F_1\left(1,\s0d2;1+\s0d2;\xi\right)
\eeq
and
\bea\label{l-wze-2}
\bar B_d(\xi) &=&
\012
\left[
\ {}_2F_1\left(1,\s0d2;1+\s0d2;\xi\right)
-
\0{d}{d+2}\,{}_2F_1\left(1,1+\s0d2;2+\s0d2;\xi\right)
\right]\,.
\eea
For $|\xi|<1$, \Eq{l-wze} can be Taylor-expanded in $\xi$, leading to 
the first equation given in \Eq{ld0Series}.

For applications, it will be useful to obtain explicit analytical
expressions for the functions \eq{l-low} for fixed dimensions. For
$d=4$, we find
\bea
\ell^4_0(\omega,z,\eta)&=&
\01{z-1}
-\0{1+\omega}{(z-1)^2}\ln\0{z+\omega}{1+\omega}
\nonumber\\ \label{l40res}
&&-\eta\left[
\0{1+2\omega+z}{4(z-1)^2}
-\012\0{(1+\omega)(\omega+z)}{(z-1)^3}
\ln\0{z+\omega}{1+\omega}
\right]
\eea 
In $d=3$ dimensions, we find
\bea
\ell^3_0(\omega,z,\eta)&=&
\0{2}{(z-1)}
-\0{2\sqrt{1+\omega}}{(z-1)^{3/2}}\arctan\sqrt{\0{z-1}{1+\omega}}
\nonumber\\ \label{l30res}
&&-\eta\left[
\0{1+3\omega+2z}{3(z-1)^2}
-\sqrt{\0{1+\omega}{z-1}}\0{z+\omega}{(z-1)^2}
  \arctan\sqrt{\0{z-1}{1+\omega}}
\right]
\eea  
and the region for $z<1$ is obtained through analytical continuation.
Finally, for $d=2$ we find
\beq \label{l20res}
\ell^2_0(\omega,z,\eta)=
\01{z-1}\ln\0{z+\omega}{1+\omega}
-\eta\left[-\012\01{z-1}+\0{z+\omega}{2(z-1)^2}
  \ln\0{z+\omega}{1+\omega}\right]
\eeq  
Notice that the functions \eq{l40res}, \eq{l30res} and \eq{l20res}
behave smoothly for $z\approx 1$, which follows either from
\Eq{ld0Series} or by an explicit check.\step

For $z=1$ these expressions reduce to the result \Eq{ld0Eta}. 

%********|*********|*********|*********|*********|*********|*********|****
\section{High temperature limit}
%********|*********|*********|*********|*********|*********|*********|****

The high-temperature limit is reached for $\tau\ge 1$, or $T\ge 2\pi
k$. Then, only the $n=0$ Matsubara mode contributes to the flow. Using
\Es{b-rt}, \eq{At0} and \eq{l-wze-1}, we find:
\beq\label{l-wzt1}
B_d(\xi,\tau\ge 1,\eta)=
\0{d}{d-1}\,\0{v_{d-1}}{v_d}\,\0{\tau}{2\pi} 
B_{d-1}(\xi,\tau=0,\eta)
\eeq
For $|\xi|<1$, \Eq{l-wzt1} can be Taylor-expanded as
\beq\label{l-wzt1-3}
B_{d}(\xi,\tau\ge 1,\eta)=
\0{d}{d-1}\,\0{v_{d-1}}{v_d}\0{\tau}{2\pi} 
\sum_{n=0}^\infty
\0{(d-1)\xi^n}{2n+d-1}
\left(1-\0{\eta}{d+1+2n}\right)\,.
\eeq
Splitting $B_{d}(\xi,\tau\ge 1,\eta)$ as in \Eq{b-factorisation}, we have
\beq
B_{d}(\xi,\tau\ge 1)=
\0{d}{d-1}\,\0{v_{d-1}}{v_d}\0{\tau}{2\pi}\,
{}_2F_1\left(1,\s0{d-1}{2};\s0{d+1}{2};\xi\right)\,.
\eeq
In turn, the flow proportional
to the anomalous dimension reduces to
\bea
\bar B_{d}(\xi,\tau\ge 1)&=&
\012\,
\0{d}{d-1}\,
\0{v_{d-1}}{v_d}
\0{\tau}{2\pi}
\times \nonumber \\ 
\label{lb-wzt1}
&&
\left[
{}_2F_1\left(1,\s0{d-1}{2},\s0{d+1}{2},\xi\right)
-
\s0{d-1}{d+1}\,
{}_2F_1\left(1,\s0{d+1}{2},\s0{d+3}{2},\xi\right)
\right]
\,.
\eea

%********|*********|*********|*********|*********|*********|*********|****

\end{document}